\documentclass[11pt, aps, pre, onecolumn, showpacs, showkeys, groupedaddress, nofootinbib]{revtex4-2}
\usepackage[utf8]{inputenc}

\usepackage[english]{babel}
\usepackage[T1]{fontenc}

\usepackage{graphicx}
\usepackage{float}
\usepackage{array}
\usepackage{dsfont}
\usepackage{epstopdf}
\usepackage{indentfirst}
\usepackage{amsmath}
\usepackage{amsfonts}
\usepackage{amssymb}
\usepackage{comment}
\usepackage[colorlinks=true, allcolors=blue]{hyperref}

\usepackage{blindtext}
\usepackage{ulem}
\usepackage{tikz}
\usepackage{soul}
\usepackage{silence}
\usepackage{color}
\usepackage{xcolor}
\date{\today}

\begin{document}

\title{Rotational Symmetry-Breaking effects in the Kuramoto model}

\author{Antonio Mihara$^1$ and Rene O.\ Medrano-T.$^{1,2}$}

\affiliation{$^1$Departamento de F\'isica, Universidade Federal de S\~ao Paulo, UNIFESP, 
09913-030, Campus Diadema, S\~ao Paulo, Brazil}

\affiliation{$^2$Departamento de F\'isica, Instituto de Geoci\^encias e Ci\^encias Exatas,
Universidade Estadual Paulista, UNESP, 13506-900, Campus Rio Claro, S\~ao Paulo, Brazil}

%%%%%%%%%%%%%%%%%%%%%%%%%%%%%%%%%%%%%%%%%%%%%%%%%%%%%%%%%%%%%%%%%%%%%%%%%%%%%%%%%%%%%%%%%%%%%%%%%%%%
%\input{Sections/2-Abstract}

\begin{abstract}
    We study the bifurcations and phase diagram for a network of identical Kuramoto oscillators with a coupling that explicitly breaks the rotational symmetry of the equations.
    Applying the Watanabe-Strogatz ansatz, the original $N$-dimensional dynamics of the network collapses to a system of dimension two. Our analytical exploration uncovers bifurcation mechanisms, including transcritical, saddle-node, heteroclinic, Hopf, and Bogdanov-Takens bifurcations, that dictate transitions between collective states. 
    Numerical validation of the full system confirms emergent phenomena such as oscillation death, global synchronization, and multicluster dynamics. By integrating reduced-model bifurcation theory with large-scale simulations, we map phase diagrams that link parameter regimes to distinct dynamical phases. 
    This work offers insights into multistability and pattern formation in coupled oscillator systems.
    Notably, the multicluster obtained exhibits behavior closely resembling the frequency-synchronized clusters identified in Hodgkin-Huxley neuron models.
\end{abstract}
\maketitle

%%%%%%%%%%%%%%%%%%%%%%%%%
%\input{Sections/3-Introduction}

\section{Introduction}
Since the seminal studies of Winfree on biological rhythms at the end of the 1960s \cite{Winfree1967}, 
phase oscillators have been playing a central role in the study of collective behavior in network dynamics.
The Kuramoto model \cite{Kuramoto1975,Kuramoto1984} is certainly the most famous derivation of the 
Winfree model and has been successfully applied to understand synchronization and related phenomena in 
several areas of science. This model and its variants are mathematically simple 
allowing theoretical insights about collective behavior stability \cite{Mihara2019,  budzinski2022geometry, budzinski2023analytical, Mihara2025, Sinha2025}, 
basin of synchronization in networks \cite{Wiley2006,Delabays2017,Mihara2022}, and control methods for 
synchronization in dense and sparse networks \cite{townsend2020dense,mihara2022sparsity}. At %} 
the same time, they can exhibit rich and complicated collective dynamics \cite{Acebron2005,rodrigues2016kuramoto}, 
having been used to study a wide variety of phenomena in complex systems: synchronization in 
%%flashing fireflies \cite{Ermentrout1991}, 
circadian rhythms \cite{Antonsen2008}, swarming dynamics \cite{Keeffe2017}, cardiac pacemaker cells \cite{Osaka2017}, 
Josephson junctions \cite{Wiesenfeld1998}, power-grid networks \cite{Dorfler2013} 
and also for earthquake sequencing \cite{Vasudevan2015}.

%%% KM c/ QSR e efeitos observados, falar sobre os trabalhos

The original Kuramoto model (KM) has rotational symmetry, i.e.\ the dynamics is invariant under rotation of 
all phase oscillators by an arbitrary angle ($\theta_j \rightarrow \theta_j+C,\,\forall j$).
This symmetry allows one to make the transformation $\theta_j(t) \rightarrow \theta_j(t) + \Omega.t,$ $\forall j, t$,
which is equivalent to viewing the dynamics in the so-called {\it comoving frame}, i.e.\ a frame rotating 
with constant angular frequency $\Omega$ with respect to the laboratory frame.
But recently the authors of Ref.\cite{ChandrasekarPRE2020} studied the effects of including a term 
in the RHS of KM that breaks explicitly its rotational symmetry. That model was derived from a set of globally 
coupled Stuart-Landau limit-cycle oscillators with conjugate feedback under 
certain restrictions. On the other hand one can show that their model can also 
be seen as a generalization of the model for a population of pulse-coupled biological 
oscillators of Ref.\cite{Ariaratnam2001}. As direct consequences of such symmetry 
breaking: (i) it is no longer possible to analyze the dynamics in the comoving frame, 
only in the laboratory frame and (ii) they observe a collective dynamical state in which 
the nodes are (roughly) in phase and stationary in the laboratory frame, a behavior that is not observed 
in the original KM.

%%% Falar sobre o modelo estudado; formalismo de WS; bifurcações e estados coletivos

In this paper we study the Kuramoto model with broken rotational symmetry for the case where the oscillators are identical, that is, all with the same natural frequency $\omega\neq 0$.
For this purpose, we  consider the Watanabe-Strogatz (WS) approach 
\cite{WS.PRL1993, WS.PhysD1994, pikovsky2010partiallyintegrabledynamicsensembles} that allows replacing the $N$ Kuramoto-like ODEs with just three ODEs for the new variables 
$z\equiv \rho e^{i\phi}$ and $\Psi$ (plus a number of constants of motion). One can show 
that, for large $N$, the variable $z$ is approximately equal to the complex Kuramoto order parameter 
\cite{pikovsky2010partiallyintegrabledynamicsensembles}. 
Then, the $N$-dimensional collective dynamic of the network is encoded in a planar phase space whose solutions and bifurcations of such model gives an idea of the possible different collective states that could be found in each region of the parameter space.
Our investigation brings to light a rich scenery of bifurcations, including transcritic, SNLC, heteroclinic, Hopf, and Bogdanov-Takens, which drastically disturbs the collective dynamics. 
A kind of roadmap, in association with numerical integrations of the $N$ Kuramoto-like ODEs, allows us to build phase diagrams and verify such different collective states of the system.

%%% Estrutura do paper
The text is organized first by introducing the model in Sec. \ref{sec2}. To uncover the 
effect of the introduced breaking rotational symmetry term, we analyze the behavior of two coupled particles in Sec. \ref{sec3}. 
Next, we move on to the $N$-dimensional network analysis. First, we obtain in Sec. \ref{sec4} the reduced system applying the 
WS approach and derive the bifurcations in the parameter space and the behavior in the phase plane.
In Sec.~\ref{sec5} we present our numerical studies through which we find a rich diversity of collective behaviors, namely,
oscillation death, synchronized states and also multicluster states.  
Finally, we present our conclusions in Sec.~\ref{sec6}.

%%%%%%%%%%%%%%%%%%%%%%%%%
%\input{Sections/4-Model}

\section{The Kuramoto model with broken rotational symmetry}
\label{sec2}

Recently it was proposed in Ref.\cite{ChandrasekarPRE2020} the addition 
of an extra interaction term to the RHS of Eq.(\ref{eq:kqsr}) that explicitly
breaks the rotational symmetry of the original KM (with $\varepsilon_2=0$):
\begin{equation}
    \dot{\theta}_j  = \omega_j + \frac{1}{N}
\left[ 
\varepsilon_1\sum_{k} \,\sin ( \theta_k - \theta_j ) +
\varepsilon_2\sum_{k\neq j} \,\sin ( \theta_k + \theta_j )
\right]
\, .
    \label{eq:kqsr}
\end{equation}

This model was derived from a set of globally coupled Stuart-Landau 
limit-cycle oscillators with conjugate feedback. However one can
observe that, for the particular case $\varepsilon_2 = - \varepsilon_1$,
the interaction term in the RHS of Eq.(\ref{eq:kqsr}) 
becomes $(-2\varepsilon_1/N) \sum_k \cos\theta_k \sin\theta_j$, the same
coupling term that appears in an idealized model for a population of 
pulse-coupled biological oscillators \cite{Ariaratnam2001}, then 
Eq.(\ref{eq:kqsr}) can be seen as a generalization of that model. 

One should notice that due to the term $\sin(\theta_k+\theta_j)$ in (\ref{eq:kqsr}), 
the transformation $\theta_j \rightarrow \theta_j + \Omega t \, (\forall\, j)$ 
does not leave the dynamics invariant in such a that there is no more equivalence
with the dynamics in the comoving frame. Therefore here we consider only the
dynamics in the laboratory frame. Without loss of generality, we also consider
only the case $\varepsilon_2 \geq 0$, since the change $\varepsilon_2 \rightarrow 
-\varepsilon_2$ is equivalent to keeping $\varepsilon_2$ unchanged but 
transforming $\theta_j \rightarrow \theta_j + \pi/2 \, , \forall\, j$ in (\ref{eq:kqsr}),
i.e.\ a redefinition of the direction with respect to which the angles are measured
\cite{ChandrasekarPRE2020}. In this work we consider the dynamical properties of the system 
(\ref{eq:kqsr}) for the case of identical oscillators, i.e.\ 
$\omega_j =\omega,\, \forall\, j$, in the framework of Watanabe-Strogatz (WS).

%%%%%%%%%%%%%%%%%%%%%%%%%
%\input{Sections/5-TwoParticles}

\section{Two oscillators behavior}
\label{sec3}

For a better perception of the actions of the system's summation terms (\ref{eq:kqsr}), it is interesting to analyze the interaction between two oscillators in the network. Then, the original model is reduced to the following two equations:

\begin{align}
        \dot{\theta}_k  &= 1 - A\,\sin ( \theta_- ) +
B\,\sin ( \theta_+ )
\, ,
    \label{eq1}
\\
        \dot{\theta}_j  &= 1 + A\,\sin ( \theta_- ) +
B\,\sin ( \theta_+ )
\, ,
    \label{eq2}
\end{align}
where $\omega_k = \omega_j \equiv\omega$, $A = \varepsilon_1/2\omega$, $B = \varepsilon_2/2\omega\geq0$, $\theta_- = \theta_k-\theta_j$ and $\theta_+ = \theta_k+\theta_j$.

\begin{figure}[ht]
    \centering
    \includegraphics[width=0.95\textwidth]{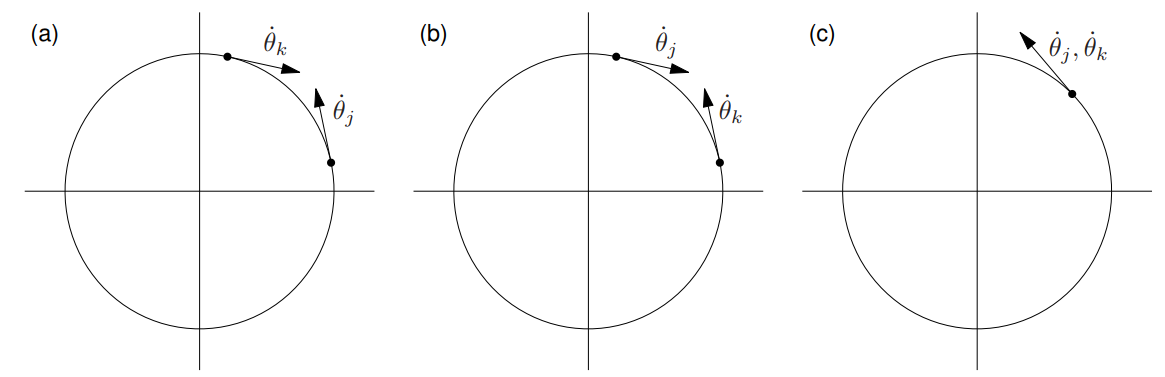}
    \caption{Attractive regime of the Kuramoto coupling for $A\gg 1$ and $B=0$ in Eqs. (\ref{eq1}) and (\ref{eq2}). The case $\theta_k > \theta_j$ (a) and $\theta_j > \theta_k$ (b) are equivalent evolutions leading both oscillators to the same phase $\theta_k = \theta_j$ (c).}
    \label{fig:1}
\end{figure}

{\bf Kuramoto Coupling Term:} Considering $B=0$ and $\omega\to 0$ (i.e., $|A| \gg 1$), Eqs. (\ref{eq1}) and (\ref{eq2}) show that the phase velocity of both oscillators are in opposite directions along the shorter path between them, if $A>0$, leading the phases to a same value as shows Fig. \ref{fig:1}.
That is the attractive regime of the Kuramoto coupling. 
If $A < 0$, the regime is called repulsive since the phases separate approaching asymptotically to $\theta_- = \pi$.
%Nevertheless, these solutions are not equilibria, they are in-phase ($A>0$) or antiphase ($A<0$) periodic solutions.

{\bf Symmetry-Breaking Coupling Term:}

\begin{figure}[ht]
    \centering
    \includegraphics[width=0.95\textwidth]{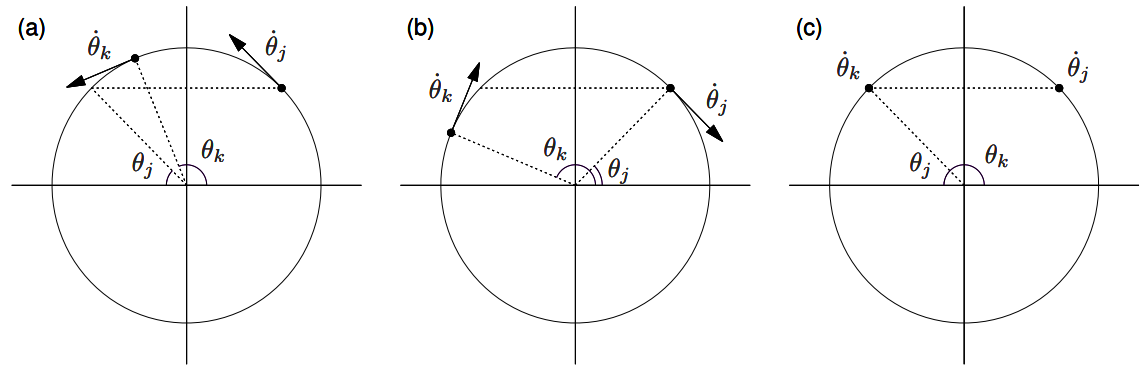}
    \caption{State of intrinsic equilibrium in coupling term without rotational symmetry ($A=0$ and $B\gg 1$). The case $\theta_k + \theta_j < \pi$ (a) and $\theta_k + \theta_j > \pi$ (b) lead both oscillators to symmetric phases around $\pi/2$ (c).}
    \label{fig:2}
\end{figure}

Observe that in the Kuramoto regimes, the system has rotational symmetry, not changing with the transformation $\theta \to \theta + C$, with $C$ constant. 
However, fixing $A=0$ and mantaining $\omega \to 0$ (now meaning, $B \gg 1$), the system loses its rotational symmetry. 
Note that, if $0<\,\theta_+\mod 2\pi<\pi$, both oscillators move counterclockwise while, for $\pi<\theta_+ \mod 2\pi<2\pi$, they move clockwise.
In both cases, the oscillator phases will move until achieve $\theta_+ = \pi$ or $\theta_+ = 3\pi$. 
The first solution is squematically presented in Fig. \ref{fig:2}. 
In (a) the oscillatores are moving counterclockwise ($\theta_+ < \pi$) whereas in (b) they are rotating clockwise ($\theta_+ > \pi$). 
Both oscillators achieve the stable equilibrium at $\theta_+ = \pi$, with phases symmetrically oposed around $\pi/2$ as shows Fig. \ref{fig:2} (c). 

The other solution, $\theta_+ = 3\pi$, results in an equilibrium in which the phases remain symmetric around $3\pi/2$.

\subsection{Stability analysis}
\label{Stablity}

%%%%%%%%%%%%%%%%%%%%%%%%%%%%%%%%%%%%%%%%%%%%%%%

In general, the behavior of the oscillators is a balance between the Kuramoto and symmetry-breaking coupling terms through parameters $A$ and $B$.  
Changing the coordinate system to $\theta_+$ and $\theta_-$ we obtain an uncoupled system indicating that these variables evolve independently:
\begin{align}
    \dot{\theta}_-  &= -2A\,\sin (\theta_-)
\, , \label{eq4}
\\
    \dot{\theta}_+  &= 2B\sin (\theta_+) +2 \, .
    \label{eq5}
\end{align}
The domains of $\theta_-$ and $\theta_+$ are given by the $\theta_k$ and $\theta_j$ phases. 
Then, while the range of $\theta_-$ is $[0,2\pi)$, the analyses of the behavior of $\theta_+$ must consider a larger one $\theta_+ \in (0,4\pi]$.

\begin{figure}[ht]
    \centering
    \includegraphics[width=0.85\textwidth]{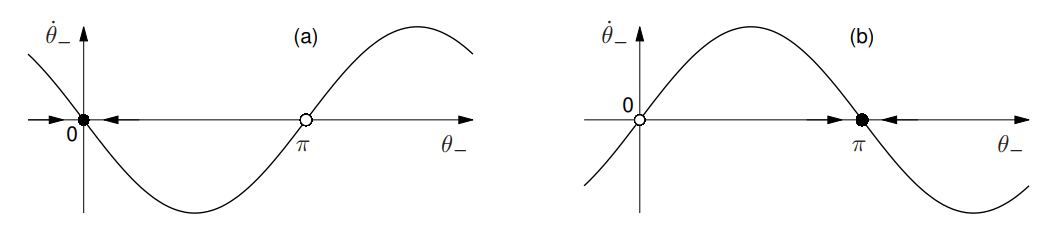}
    \caption{States due to Eq. (\ref{eq4}). (a) Attractive regime for $A>0$. (b) Repulsive regime for $A<0$. The arrows indicate the direction that moves $\theta_-$. Black and white dots denote stable and unstable states, respectively.}
    \label{fig:Q-}
\end{figure}

Figure \ref{fig:Q-} shows that Eq. (\ref{eq4}) has two equilibria where the oscillators are in phase ($\theta_- = 0$) and antiphase ($\theta_- = \pi$), essential conditions for the attractive and repulsive Kuramoto regimes. 
For $A>0$ the in-phase state is stable while the antiphase is unstable [Fig. \ref{fig:Q-}(a)]. 
For $A<0$ the stability is exchanged 
[Fig. \ref{fig:Q-}(b)]. Note that solutions, given by Eq. (\ref{eq4}), do not impose specific values for the oscillator phases.

\begin{figure}[ht]
    \centering
    \includegraphics[width=0.85\textwidth]{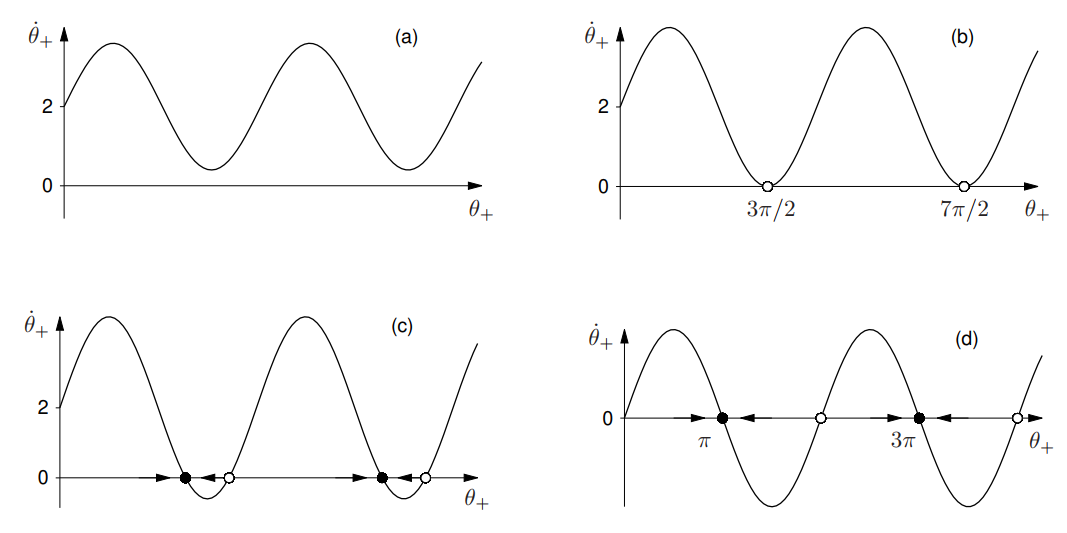}
    \caption{Equilibrium states for Eq. (\ref{eq5}). Arrows indicate the direction of $\theta_+$'s evolution, while black and white dots denote stable and unstable equilibria, respectively. (a) $0<B<1$: The system exhibits no equilibrium points. (b)  $B = 1$: A tangent bifurcation occurs. (c) $B > 1$: Stable and unstable equilibrium points bifurcate and separate for moderate values of $B$. (d) $B \gg 1$ with $\omega \to 0$: The stable and unstable equilibrium states are separated by a phase difference of $\pi$.}
    \label{fig4}
\end{figure}

Figure \ref{fig4} presents important dynamical aspects of $\theta_+$. According to Eq. (\ref{eq5}).
In (a) if $0\leq B<1$, $\dot{\theta}_+$ is never null, then the phase oscillators run counterclockwise around the circle periodically. 
In (b) a {\it saddle-node} bifurcation takes place when $B = 1$ giving rise to two unstable states at $\theta_+ = 3\pi/2$ and $\theta_+ = 7\pi/2$. 
In (c), as $B>1$ increases, a pair of stable states moves in the range $\, \theta_+^{S_1} \in [\, \pi, 3\pi/2 \,)$ and $\, \theta_+^{S_2} \in [\, 3\pi, 7\pi/2\, )$. 
As discussed before, this state is achieved when $\theta_k$ and $\theta_j$ are symmetrically opposed around $\pi/2$ as shown in Fig. \ref{fig:2} (c) for $A=0$ and $B \gg 1$.
The unstable states are $\, \theta_+^{U_1} \in [\, 3\pi/2, 2\pi\, ]$ and $\, \theta_+^{U_2} \in [\, 7\pi/2, 4\pi \,]$ as shows Fig. \ref{fig4} (d).

Considering all the states supplied independently by each Eqs. (\ref{eq4}) and (\ref{eq5}), the parameter space can be split into four regions with different dynamical behavior as follows.

\begin{figure}[ht]
    \centering        \includegraphics[width=0.5\textwidth]{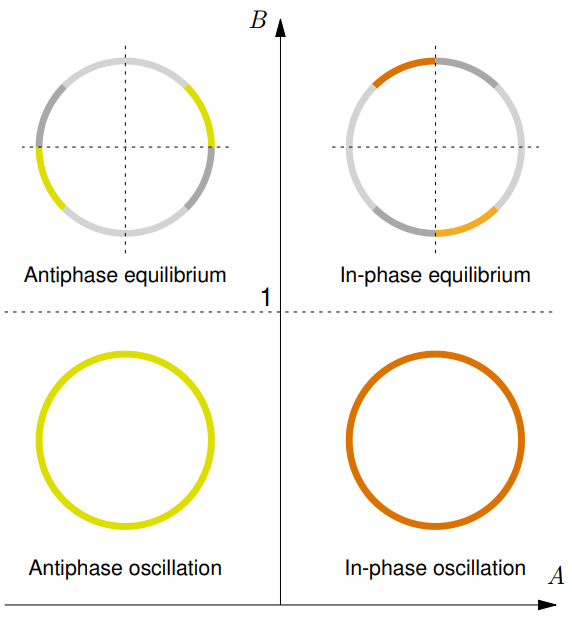}
    \caption{Dynamical states of the interaction between two phase oscillators in the unitary circle. The states with $\theta_k = \theta_j$ (in-phase) and $\theta_k - \theta_j = \pi$ (antiphase) are colored in orange and yellow, respectively. For $B>1$, the light and dark orange indicate the allowed region for two coexistent equilibrium while the dark gray corresponds to repeller and saddle types of equilibria. The circles monotonically colored represent periodic oscillations.}
    \label{fig5}
\end{figure}

{\bf $A>0$ and $0<B<1$:} 
In this case, $\dot{\theta}_+ \neq 0$ and $\theta_- = 0$ is stable while $\theta_- = \pi$ is unstable. 
For these conditions, the phases $\theta_k$ and $\theta_j$ collapse in a same phase and oscillate periodically with velocity given by Eq. \ref{eq5}. 
That is the {\it in-phase oscillation} represented in Fig. \ref{fig5} by a filled circle. 
The color orange indicates that $\theta_k = \theta_j$ is stable. 
The periodic oscillation in antiphase $\theta_k - \theta_j = \pi$ is also possible but it is unstable.

{\bf $A<0$ and $0<B<1$:} 
For $A<0$, the stability states are reversed of the latter case, $\theta_- = 0$ is unstable, while $\theta_- = \pi$ is stable. 
Thus, the stable periodic behavior has the phases $\theta_k$ and $\theta_j$ $\pi$ distant, and the unstable behavior is in phase. 
In Fig. \ref{fig5}, the stable {\it antiphase oscillation} is represented by the yellow collor.

{\bf $B>1$:} In this case, $\theta_k$ and $\theta_j$ should remain symmetrically posed around $\theta_+/2$. 
However, if $\theta_-$ is in the attractive regime ($A>0$), the oscillator phases join in a stable equilibrium at $\theta_k = \theta_j = \theta_+^{S_i}/2$. There are two equilibria, one for $i=1$, in the range $( \,\pi/2,\,3\pi/4 \, )$ and another, for $i=2$,  in the range $( \, 3\pi/2, \,7\pi/4 \,)$.
We call these stationary states as {\it in-phase equilibrium}. They are depicted by the dark and light orange color, respectively, in Fig. \ref{fig5}.
Contrary, if $A<0$, an {\it antiphase equilibria} becomes stable. 
Considering conditions $\, \theta_k - \theta_j = \pi$ and $\, \theta_k + \theta_j = \theta_+^{S_i}$ this equilibrium is composed by $\, \theta_k \in ( \, \pi, 5\pi/4 \,)$ and $\, \theta_j \in ( \, 0, \pi/4 \,)$ for $i=1$. 
For $i=2$, $\theta_k$ and $\theta_j$ exchange their phase. 
Since the oscillators are indistinguishable, we consider the same solution.
The yellow strips in Fig. \ref{fig5} indicate the regions of the antiphase equilibrium.
Note that when the antiphase equilibrium is stable, the in-phase state becomes a saddle-type equilibrium and vice versa. 
Furthermore, considering the unstable state $\theta_+^{U_i}$ with the unstable state of $\theta_-$ [namely $\theta_- = \pi$ if $A>0$ and $\theta_- = 0$ if ($A<0$)] three repulsive equilibria arise: one antiphase in $[\, \pi/4, \pi/2 \,]$ and $[\, 5\pi/4, 3\pi/2 \,]$, the other two in phase in $[\, 3\pi/4, \pi \,]$ and $[\, 7\pi/4, 2\pi \,]$. 
The saddle and repeller regions are shown in Fig. \ref{fig5} in light and dark gray, respectively. 
For $B<0$, the behavior is equivalent.

Regarding the bifurcations of the system, when $B = 1$, two supercritical saddle-node bifurcations occur for $A > 0$ ($\theta_k=\theta_j$), in which two equilibria (sink and saddle) emerge at $3\pi/4$ and another two at $7\pi/4$. These equilibria are also interconnected, forming a heteroclinic orbit. 
Consequently, this constitutes a {\it saddle-node heteroclinic} bifurcation. 
For negative values of $A$, the heteroclinic orbit similarly arises through a subcritical saddle-node bifurcation (source and saddle).
Equivalent behavior is observed for the equilibrium $\theta_k-\theta_j=\pi$.
Such bifurcations can also be seen as {\it saddle-node bifurcations on a limit cycle} (SNLC).
Due to the periodicity of the variables $\theta_{\pm}$, one can realize that
their state space is a torus. By direct analysis of Fig. \ref{fig:Q-} and
Fig. \ref{fig4}(a) one can infer that for $0<B<1$ there is a 
stable limit cycle in $\theta_{-}=0\, (\pi)$ if $A>0\, (A<0)$.
But for $B=1$ two fixed points appear on each cycle $\theta_{-}=0,\, \pi$
(as shown in Fig.\ref{fig4}(b)).
For $B>1$ each fixed point splits into a saddle and a node, but with all of them on the cycles.

A notable degeneracy occurs when $A = 0$, as any difference $ \theta_k - \theta_j $ becomes permissible.
When $0< B < 1$, multiple limit cycles emerge, a characteristic feature of a {\it degenerate Hopf} bifurcation. Conversely, for $B > 1$, additional equilibria appear, symmetrically positioned around $\pi/2$ and $3\pi/2$ (see Fig. \ref{fig:2} (c)), corresponding to $\theta_k$ and $\theta_j$ arranged in opposition.

%%%%%%%%%%%%%%%%%%%%%%%%%
%\input{Sections/6-Multiple}

\section{Multi-oscillators analysis}
\label{sec4}

%%%%%%%%%%%%%%%%%%%%%%%%%
%\input{Sections/6.1-Multiple/i-WS}

\subsection{Analysis in the WS framework (Reduced Model)}
\label{WS}

In their seminal publications Watanabe and Strogatz \cite{WS.PRL1993,WS.PhysD1994} 
showed that the dynamics of a class of systems with $N$ identical 
coupled oscillators can be described by three global variables 
$\rho, \phi, \Psi$ and $N-3$ constants of motion. The variable 
$0 \leq \rho \leq 1$ is roughly similar to the amplitude of the 
Kuramoto order parameter, while $\phi$ and $\Psi$ are angular variables. 
In general one combines two of the variables as $z = \rho e^{i\phi}$.

The Kuramoto order parameter is the complex mean field 
given by
\begin{equation}
    Z(t) = \frac{1}{N} \sum_k e^{i \theta_k (t)} \, = R.e^{i\Theta} \, ,
\end{equation}
and one can show that system (\ref{eq:kqsr}) with identical oscillators
can be approximated by:
\begin{equation}
\dot{\theta}_j  \simeq \omega + \mbox{Im}\left[ 
\left(\varepsilon_1 Z - \varepsilon_2 Z^* \right) \, e^{-i\theta_j}
\right] \, .
\label{eq:kqsr2}
\end{equation}

In WS theory the system of $N$ equations (\ref{eq:kqsr2}) can be replaced
by the following pair of ODEs for the new global variables $z, \Psi$
(and also regarding $Z \approx z$)
\begin{eqnarray}
    \dot{z} &=& i\omega z + 
    \frac{\varepsilon_1}{2} (1 - |z|^2)z +
    \frac{\varepsilon_2}{2} (z^3 -z^* )\, , \label{eqZ}\\
      \dot{\Psi} &=& \omega +
      \mbox{Im}[ \varepsilon_1 |z|^2 - \varepsilon_2 (z^*)^2 ] \, .
\label{eqPsi}
\end{eqnarray}

Formally $z$ and $Z$ are different objects, but the Kuramoto order parameter 
$Z$ is equal to $z\,\gamma(\rho,\Psi)$, with $\gamma \rightarrow 1$ for 
$N\rightarrow\infty$, for further details see Ref.\cite{pikovsky2010partiallyintegrabledynamicsensembles}.
One can also show that if $\rho=|z|=0\,(1)$ one obtains $R=|Z|=0\,(1)$. 
However for $0 < \rho < 1$ the relation between $R$ and $\rho$ is not so trivial and
depends on the variable $\Psi$ and the constants of motion. 
Furthermore, one can notice that the first ODE (\ref{eqZ}) does not depend on the variable $\Psi$, so we
can focus on studying the first equation, i.e., the dynamics in the $z$ plane,
which can provide us with some information about the collective states.
Let us divide both sides of Eq.(\ref{eqZ})
by $\omega$ (without loss of generality we suppose here that $\omega>0)$ 
and rescaling the time variable $t \rightarrow \omega t$, then we have
\begin{equation}
\dot{z} = i z + 
    a (1 - |z|^2)z + b(z^3 -z^* )\, , 
\label{eqZ2}    
\end{equation}
where $a\equiv \varepsilon_1/(2\omega)$ and $b\equiv \varepsilon_2/(2\omega)$.

%%%%%%%%%%%%%%%%%%%%%%%%%
%\input{Sections/6.1-Multiple/ii-rho=0}

\subsection{Solutions with \texorpdfstring{$\rho=0$}{TEXT}}

Now in order to analyze solutions with $\rho = |z| = 0$ of the system we shall 
rewrite Eq.(\ref{eqZ2}) above in terms of $x\equiv\mbox{Re}(z)$ and 
$y\equiv\mbox{Im}(z)$. 
\begin{eqnarray}
    \dot{x} &=& - y + (a-b)(x-x^3) - (a+3b)xy^2 \, , \nonumber\\
    \dot{y} &=& x + (a+b)(y-y^3) - (a-3b)x^2 y \, .
\label{eqXY}
\end{eqnarray}

The linear stability analysis of Eq.(\ref{eqXY}) around the equilibrium at the origin of
$xy$ plane will provide us this pair of eigenvalues
\begin{equation}
    \lambda_{\pm} = a \pm \sqrt{b^2 - 1} \, ,
    \label{lambda}
\end{equation}
which will allow us to get some bifurcation curves in the
plane of parameters.

%\subsubsection*{Pitchfork bifurcation}

By taking $\lambda_{-} = 0$, we obtain the curve 
$b=\sqrt{a^2+1}$ where occurs a {\it pitchfork bifurcation}:
for $b>\sqrt{a^2+1}$ the origin is a saddle,
and for $1<b\lesssim\sqrt{a^2+1}$ the origin
is now a %\Xout{spiral} 
sink (for $a<0$) or a %\Xout{spiral}
source (for $a>0$) and there is a pair
of saddles very close to it.

%\subsubsection*{Degenerate Hopf bifurcation}

\begin{figure}
    \centering
    \includegraphics[width=0.32\textwidth]{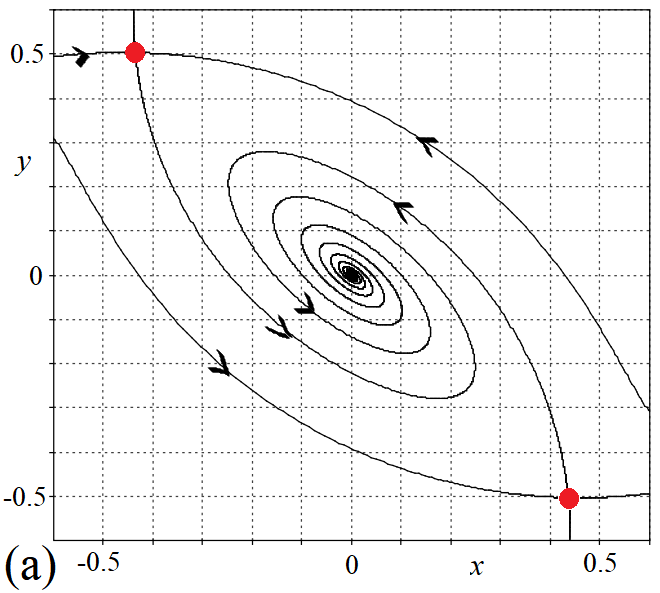}
    \includegraphics[width=0.32\textwidth]{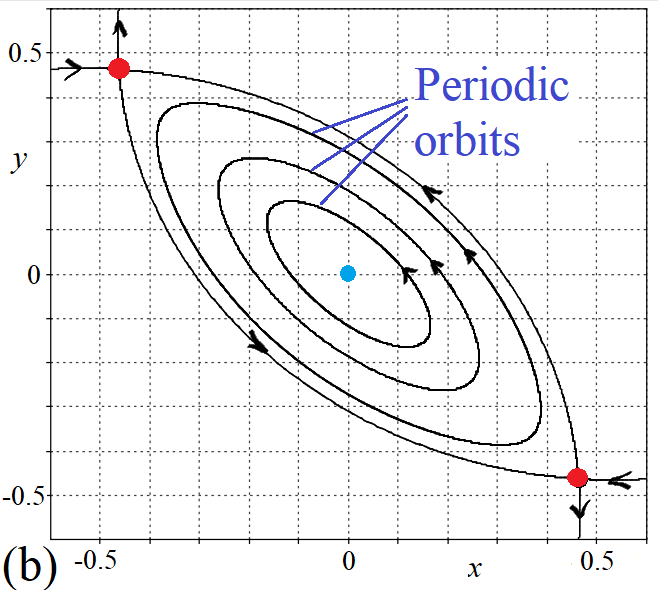}    
    \includegraphics[width=0.32\textwidth]{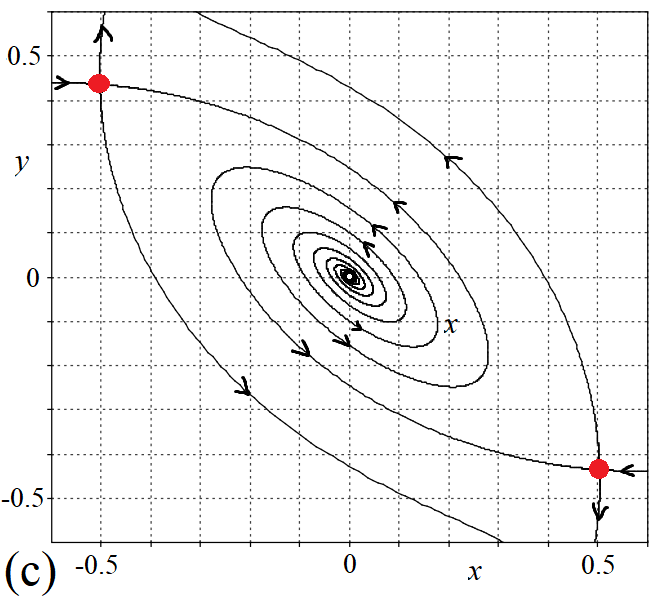}
    \caption{Heteroclinic and (degenerate) Hopf bifurcation in phase plane
    around the origin, for $b=0.7$ and: (a) $a=-0.1$ the origin is a spiral sink; (b) $a=0$ the origin is a center (blue symbol)
    surrounded by a heteroclinic loop formed 
    by the invariant manifolds of the pair of saddles (red symbols); (c) $a=0.1$ the origin is a spiral source.
    See also Fig. \ref{fig:1S} for $b = 0.68$.}
    \label{fig:hopf}
\end{figure}

For $a=0$ and $|b|< 1$, we have $\lambda_{\pm}=
\pm i\,\sqrt{1-b^2}$, the origin becomes a nonlinear center. It is a {\it Andronov-Hopf bifurcation} with 
the first Lyapunov coefficient equals to zero, see Fig.\ref{fig:hopf}.

For $a=0$ and $b=1$ both eigenvalues $\lambda_{\pm} = 0$, indicating a 
{\it Bogdanov-Takens bifurcation}, 
which is confirmed by the softwares AUTO and 
MatCont.
Numerical studies also indicate that 
{\it heteroclinic bifurcation} occurs 
for $a=0$ and $1/2 < b < 1$ : in phase
plane the origin is surrounded by a (heteroclinic) loop formed by 
the invariant manifolds of the
pair of saddles (born of the pitchfork bifurcation)
and inside the loop there is an infinite number of periodic orbits
around the origin, Fig.\ref{fig:hopf}(b). 
The eigenvalues Eq. (\ref{lambda}) also shows that only nodes can occur for $b>1$ since no complex solution is possible and spiral 
configurations are forbidden as shows Fig. \ref{fig:1S} for $b = 1.01$.

%%%%%%%%%%%%%%%%%%%%%%%%%
%\input{Sections/6.1-Multiple/iii-0<pho<1}

\subsection{\texorpdfstring{$0<\rho<1$}{TEXT}}
\label{0<rho<1}

In order to investigate other types of solutions (with $\rho>0$) we
also consider the complex feature of $z=\rho\, e^{i\phi}$ in Eq. (\ref{eqZ2})
and we rewrite the complex ODE in polar coordinates as
\begin{eqnarray}
    \dot{\rho} &=& \rho \, [a-b\cos(2\phi)] \, (1-\rho^2) \label{eq:1S}\\
    \dot{\phi} &=& 1 + b\sin(2\phi) (1+\rho^2)\, .
    \label{eq:2S}
\end{eqnarray}
Regular behavior can be analyzed, for $\dot{\rho}=0$, by its Jacobian matrix
\begin{equation}
J = 
\begin{pmatrix}
    (a-b\cos(2\phi)(1-3\rho^2) & 
    \quad 2\rho\,  b\sin(2\phi)(1-\rho^2)\\
    2\rho\, b\sin(2\phi) & 
    \quad 2(1+\rho^2)\, b\cos(2\phi)
\end{pmatrix}_{(\rho,\phi)}.
\label{eq:3S}
\end{equation}

Equation (\ref{eq:1S}) indicates three different possibilities of regular motion in terms of patterns of synchronization. 
For this analysis, we consider $b>0$ since for $b<0$ the study presents equivalent results.

Despite in the Kuramoto model $\rho > 1$ is not allowed, Eqs. (\ref{eq:1S}) and (\ref{eq:2S}) present no restrictions. 
So, for the better understanding the bifurcations in the system we consider here $\rho\geq 0$ without restrictions. 
For the equilibrium points, $(\dot{\rho}, \dot{\phi})=(0,0)$ is given if $[a-b\cos(2\phi)] = 0$ in Eq. (\ref{eq:1S}) 
and $\sin(2\phi) < 0$ in Eq. (\ref{eq:2S}).
These conditions lead to $b\cos(2\phi) = a$ and $b\sin(2\phi) = -\sqrt{b^2-a^2}$.
Therefore, for $a > 0$, there are two symmetric equilibria $(0, \phi_2^+)$ and $(0, \phi_2^++\pi)$, with $\phi_2^+ = [-\pi/4, 0)$. 
Equivalently, for $a<0$, there are two equilibria $(0, \phi_2^-)$ and $(0, \phi_2^-+\pi)$, with $\phi_2^- = (-\pi/2, -\pi/4]$. 
%These results are similar to those in subsection \ref{rho0}.
%The difference is that for $\rho \neq 0$ the equilibria are separated in the cartesian coordinates. 
The Jacobian (\ref{eq:3S}) for these equilibria is  
\begin{equation}
J = 
\begin{pmatrix}
    0 & \quad -2\, \frac{\rho(1-\rho^2)}{1+\rho^2} \\
    -2\, \frac{\rho}{1+\rho^2} & \quad 2a(1+\rho^2)
\end{pmatrix}_{(\rho, \phi^{\pm})}
\label{eq:9S}
\end{equation}
with eigenvalues
\begin{equation}
    \lambda_{1,2} = a(1+\rho^2)\left[1 \pm \sqrt{1+4\, \frac{\rho^2(1-\rho^2)}{a^2(1+\rho^2)^4}}\right]\, ,
    \label{eq:10S}
\end{equation}
where the sign of $a$ determines the pair of equilibria we are leading.
For $0<\rho<1$ the root square in Eq. (\ref{eq:10S}) is greater than one.
Therefore, $\lambda_1 \lambda_2 < 0$ and the pair of symetric equilibria are saddle equilibrium points: $(\rho, \phi_2^+)$ 
and $(\rho, \phi_2^++\pi)$, for $a>0$, and $(\rho, \phi_2^-)$ and $(\rho, \phi_2^-+\pi)$, when $a<0$.

For $\rho = 0$, $\lambda_1 = 2a$ and $\lambda_2 = 0$. To understand this bifurcation, 
observe that from Eq.(\ref{eq:2S}) an equilibrium has 
\begin{equation}
    \rho^2 = \frac{1}{\sqrt{b^2-a^2}}-1\, .
    \label{eq:11S}
\end{equation}
Taken it in account, $\rho = 0$ implies in $b = \sqrt{a^2+1}$ which is the bifurcation obtained in the last section. 
It is about the {\it pitchfork} bifurcation where an equilibrium change its stability while two equilibria born.
This scenario is presented in Fig. \ref{fig:1S} for $b = 1.01$. 
For $a=0$ at the origin there is a saddle equilibrium point that is stable (unstable) for $a=-0.5$ ($a=0.5$) surrounded 
by two symmetric saddle equilibrium points, $(\rho,\phi_2^-)$ and $(\rho,\phi_2^-+\pi)$ ($(\rho,\phi_2^+)$ and $(\rho,\phi_2^++\pi)$).

In view of the bifurcation clarity, let us consider $\rho > 1$.
It allows sinks ($\lambda_{1,2}<0$ for $a<0$) and sources ($\lambda_{1,2}>0$ for $a>0$) since the root square of 
Eq.(\ref{eq:10S}) is lesser than one (See Fig. \ref{fig:1S} for $a \neq 0$ and $b=0.68$). 
The changing of the saddle's stabilities occur for $\rho = 1$ as it gives $\lambda_1 = 4a$ and $\lambda_2 = 0$. 
Moreover, $\rho =1$ implies $b = \sqrt{a^2+1/4}$ according Eq. (\ref{eq:11S}).
We clarify this bifurcation in the next section. 
Here, the last investigation $a = 0$ results $\lambda_{1,2}=\pm\frac{2\rho}{\rho^2+1}\sqrt{1-\rho^2}$. 
Regarding the Kuramoto model ($\rho \leq 1$) and the pitchfork bifurcation ($\rho = 0$), for $a=0$ and  
$1/2 \leq b \leq 1$ the equilibria are saddles. 
Additionally, for $a=0$, $\phi_{2}^+ = \phi_{2}^- = -\pi/4$ which can be verified in Fig. \ref{fig:1S} 
with $a=0$ and $b = 0.68$. 
In this phase space local, due to the degenerated Andronov Hopf bifurcation of the equilibrium at the origin, a 
heteroclinic orbit is formed connecting the pair of saddle equilibrium points. 
Therefore, the set $(a, b) = (0, 1)$ is a {\it Bogdanov-Takens} bifurcation wich concentrates the Andronov-Hopf, 
pitchfork, and heteroclinic bifurcations.

%%%%%%%%%%%%%%%%%%%%%%%%%
%\input{Sections/6.1-Multiple/iv-rho=1}

\subsection{\texorpdfstring{$\rho=1$}{TEXT}}
\label{rho1}

\begin{figure}[hbt]
    \centering
    \hspace{1.5cm} $a = -0.5$ \hspace{2cm} $a = 0$ \hspace{2.3cm} $a=0.5$ \\
    $b=1.010$
    \includegraphics[width=0.2\textwidth]{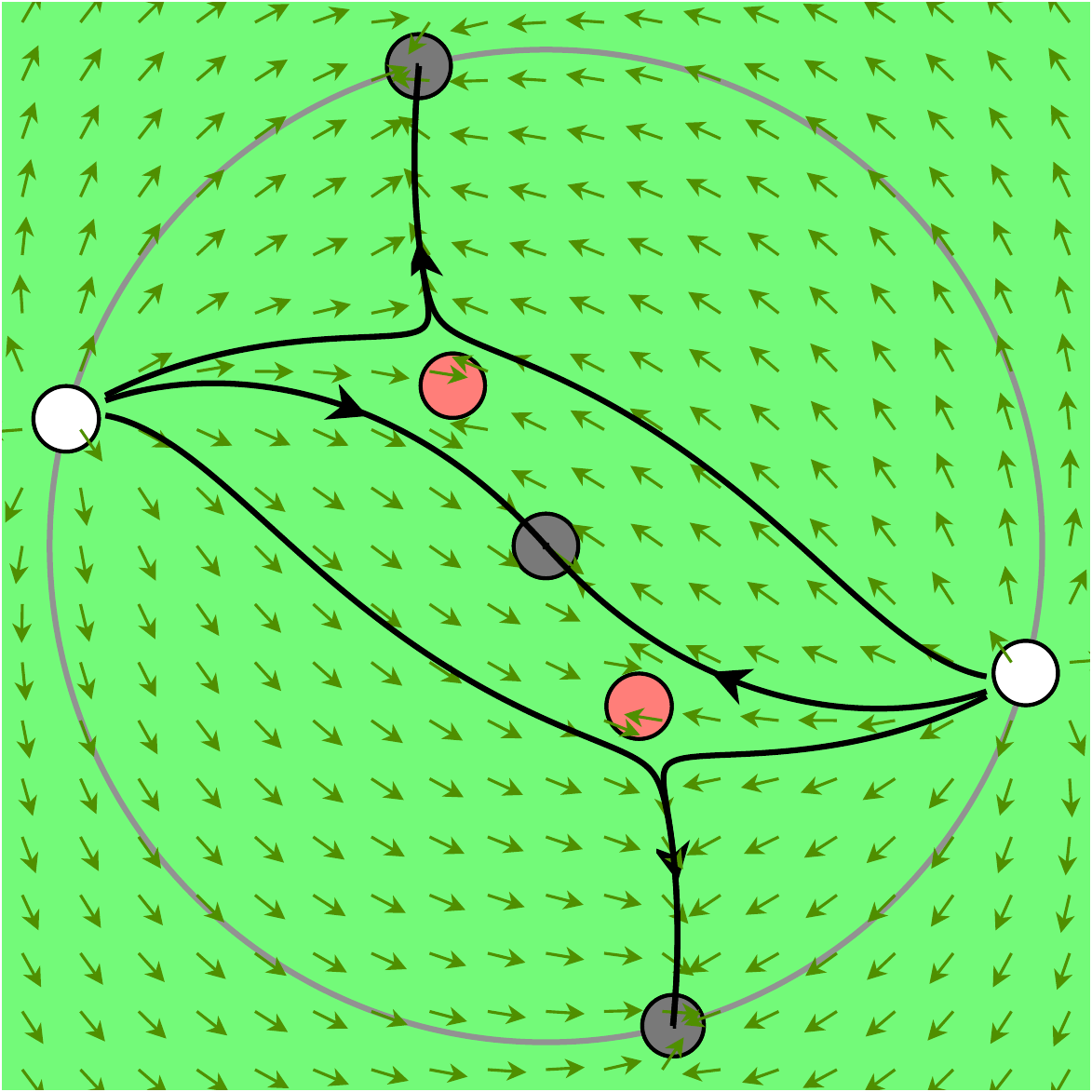}
    \includegraphics[width=0.2\textwidth]{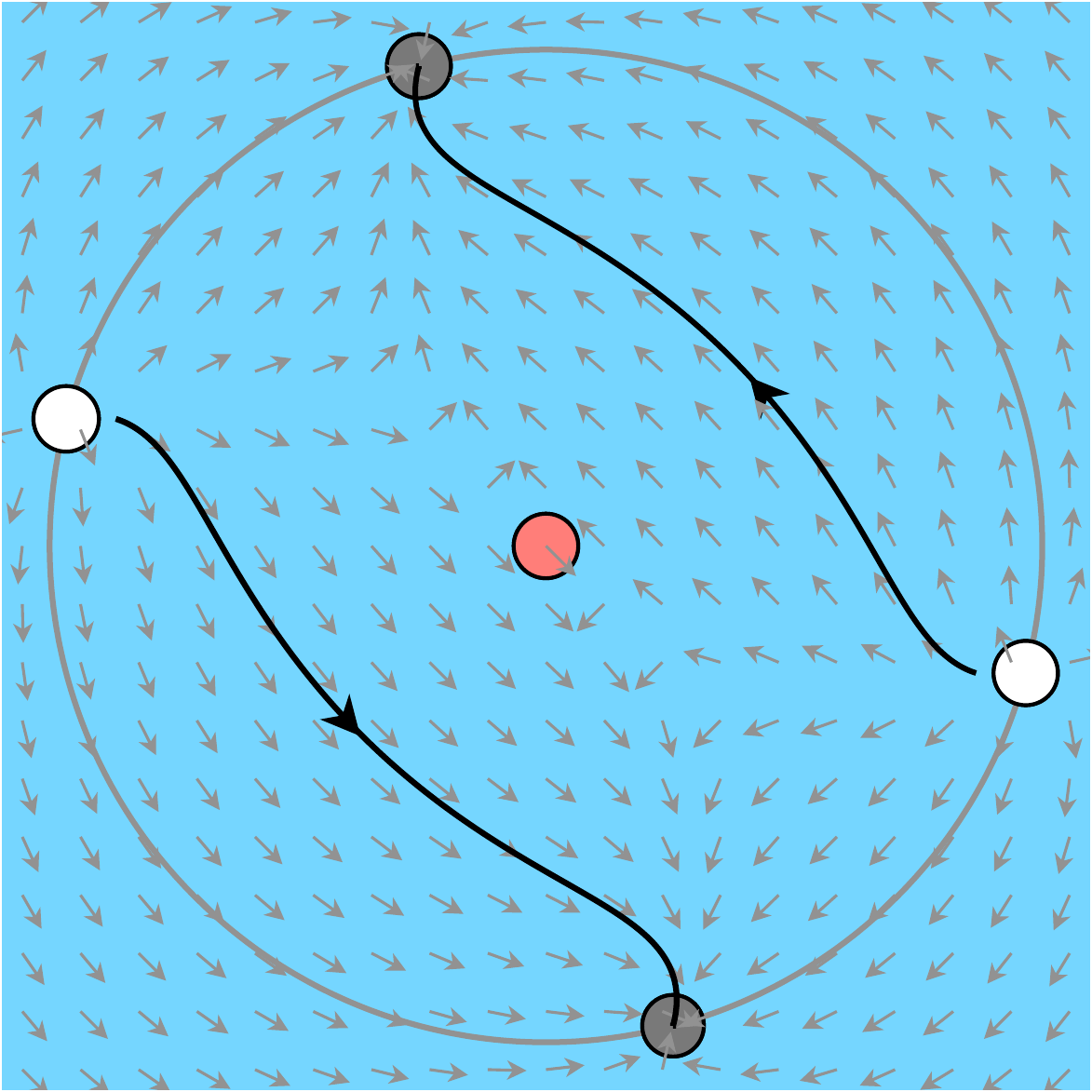}
    \includegraphics[width=0.2\textwidth]{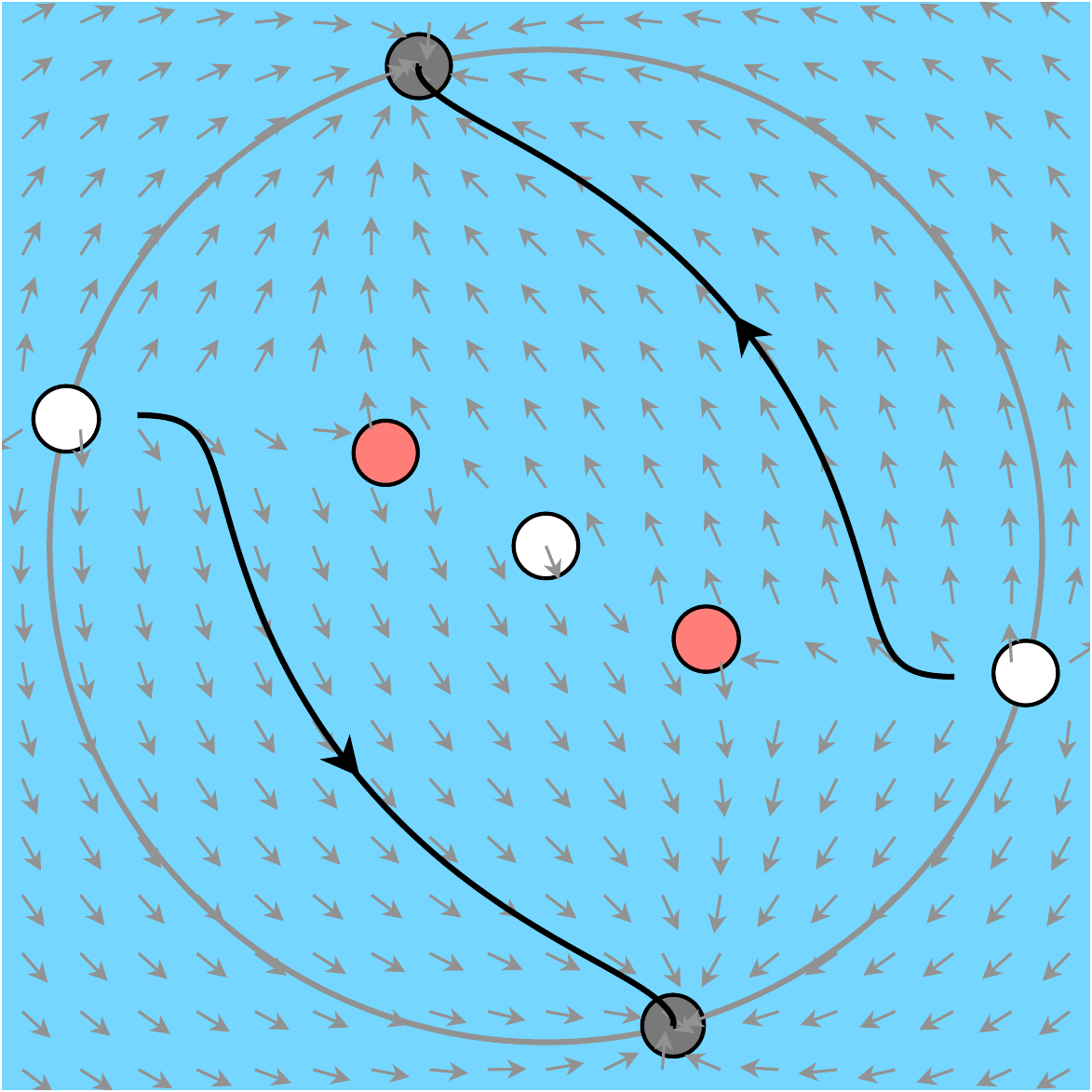} \\ \vspace{0.1cm}
    $b=0.680$
    \includegraphics[width=0.2\textwidth]{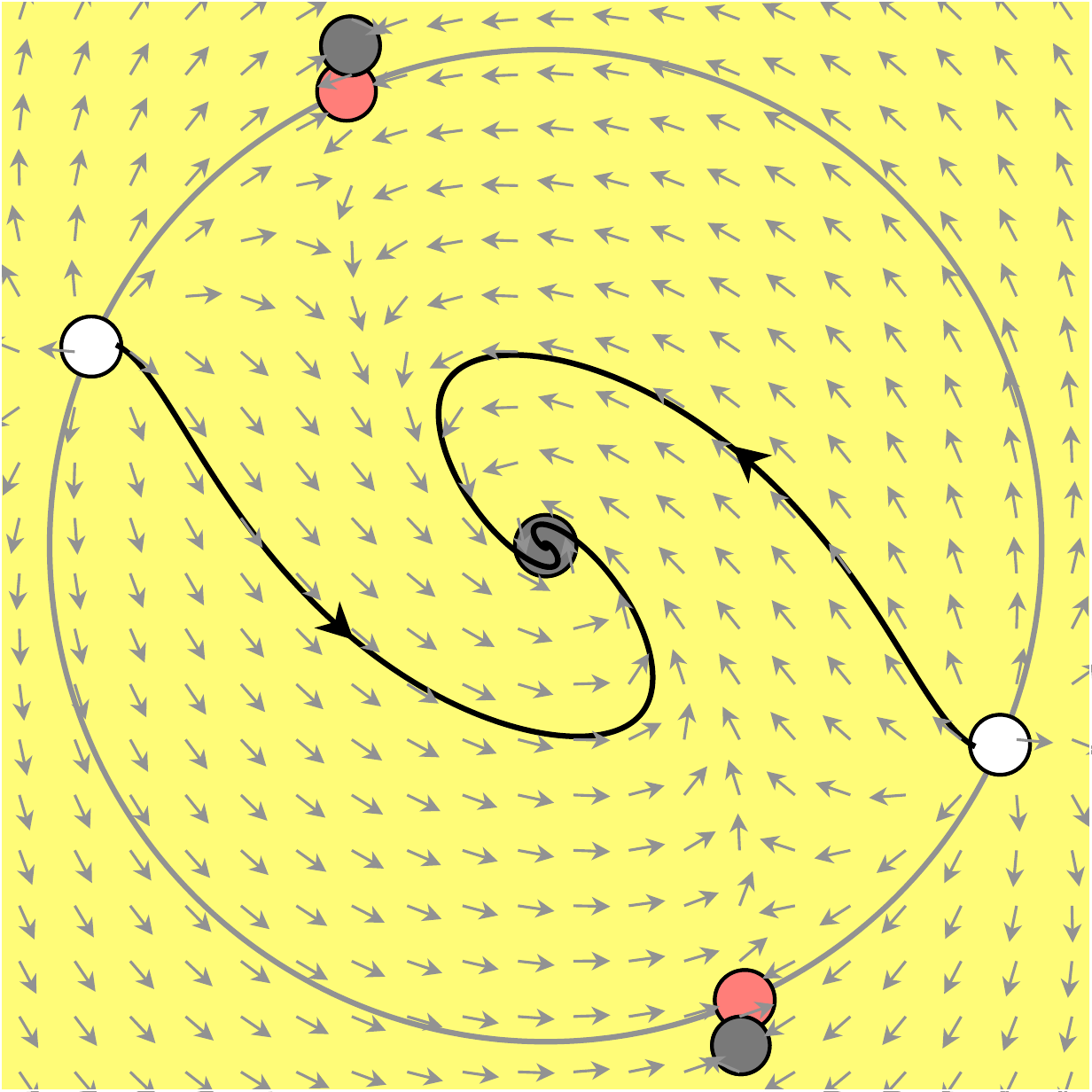}
    \includegraphics[width=0.2\textwidth]{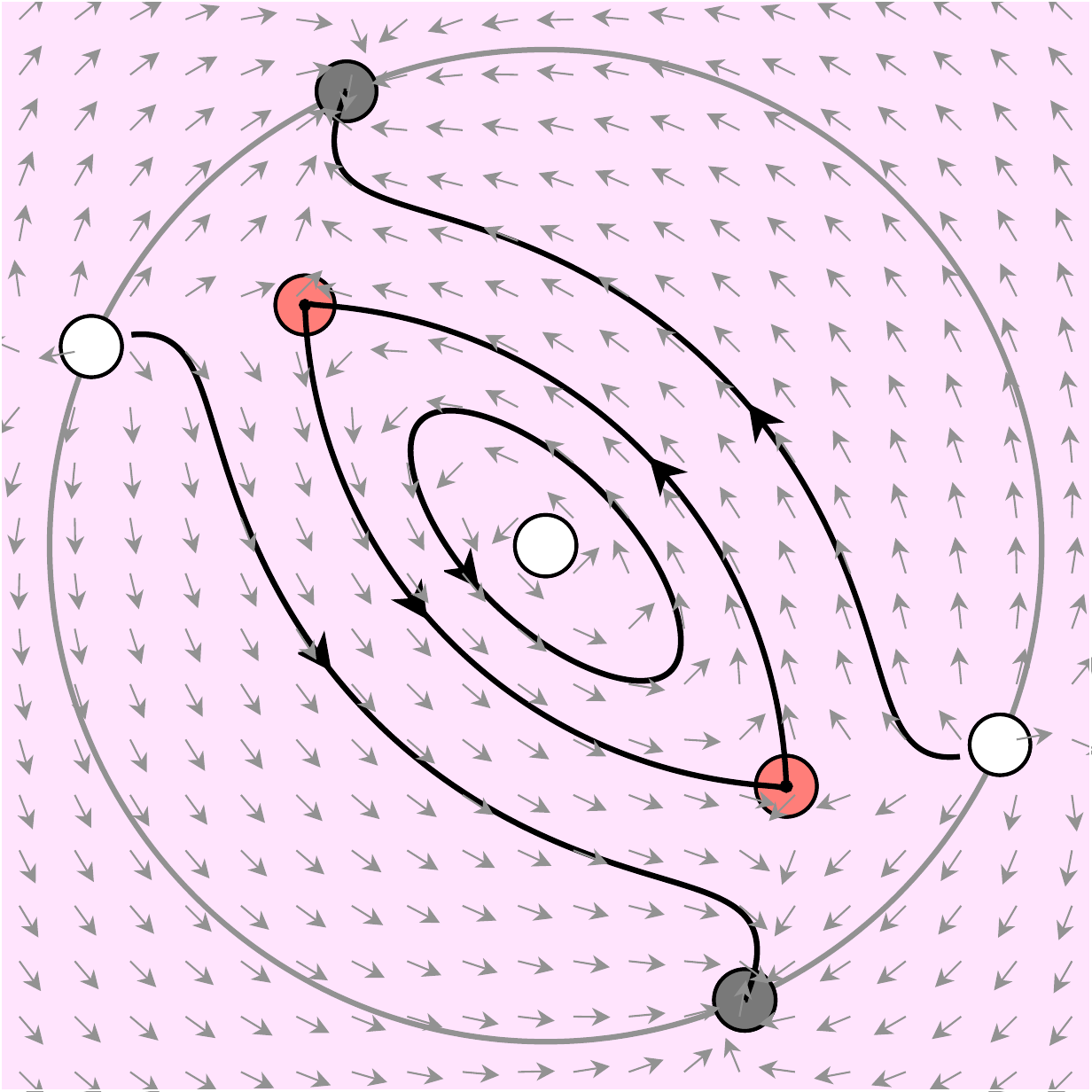}
    \includegraphics[width=0.2\textwidth]{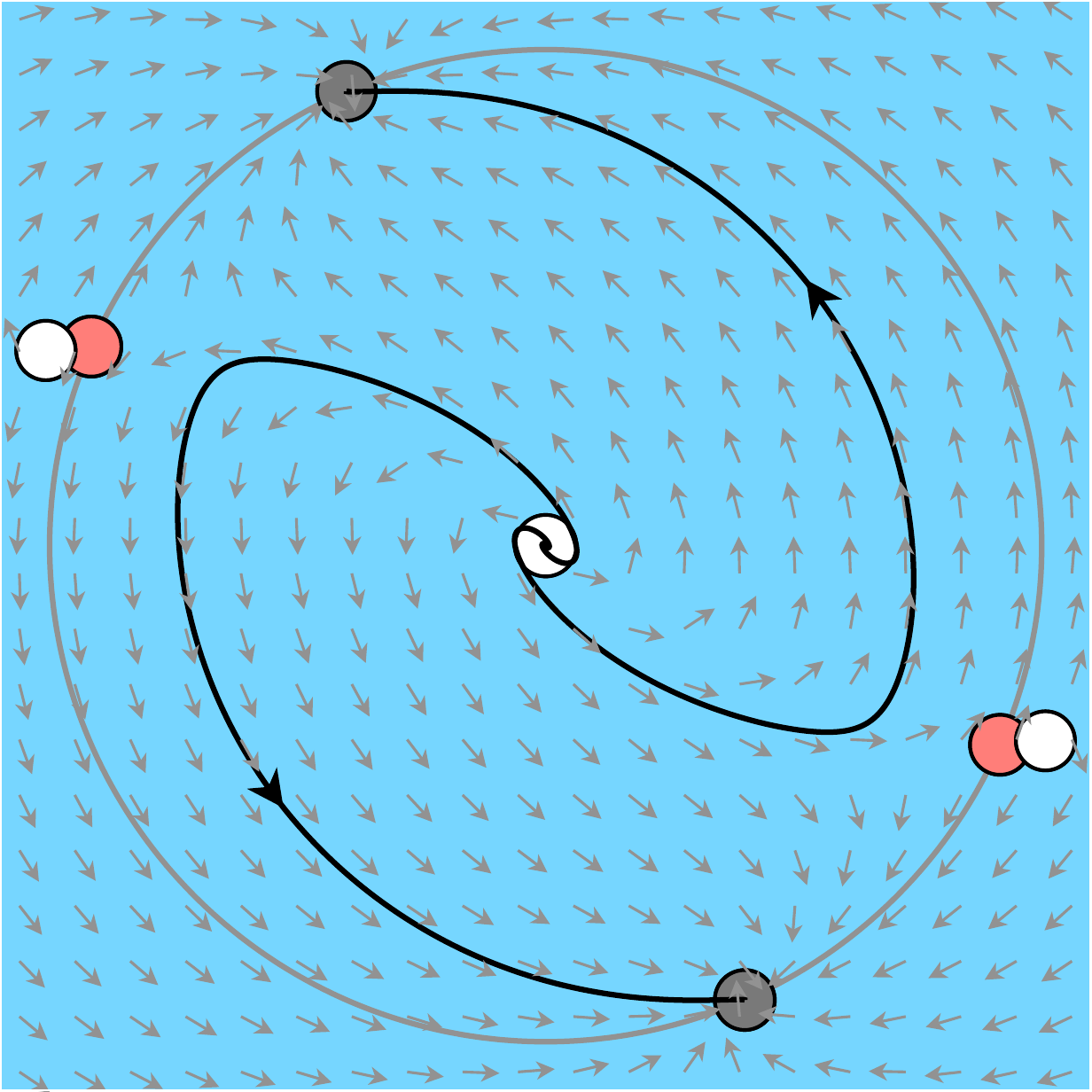} \\ \vspace{0.1cm}
    $b=0.503$
    \includegraphics[width=0.2\textwidth]{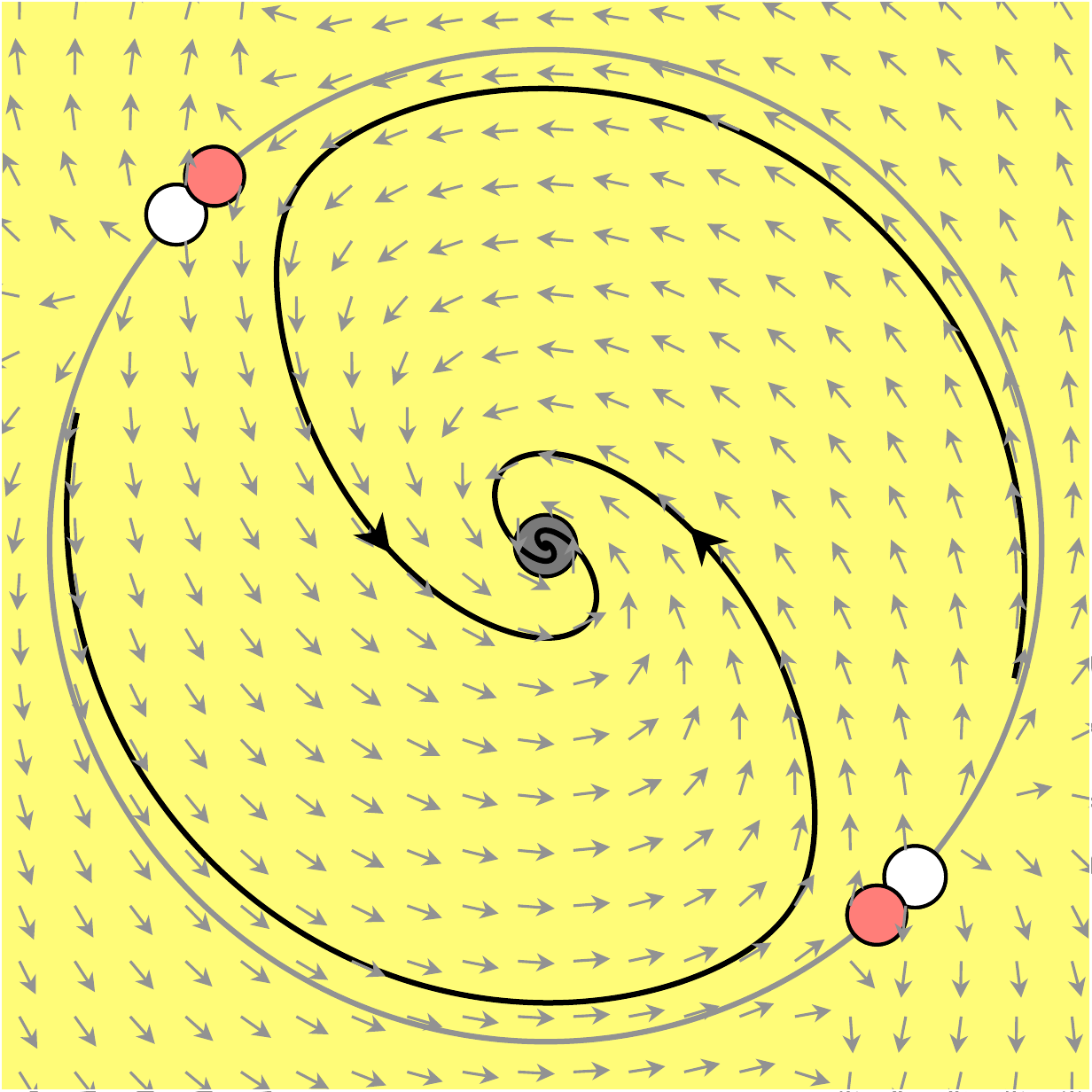}
    \includegraphics[width=0.2\textwidth]{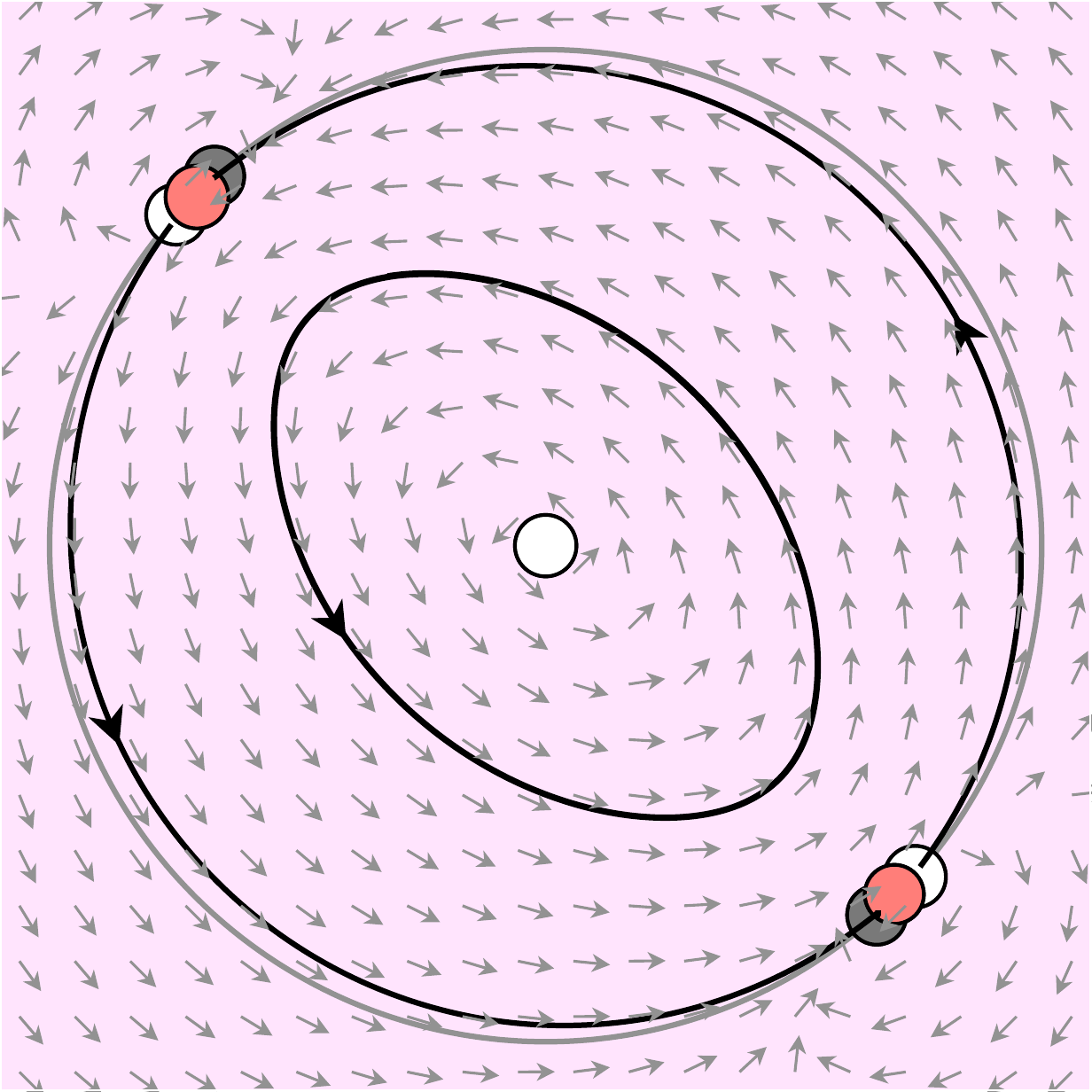}
    \includegraphics[width=0.2\textwidth]{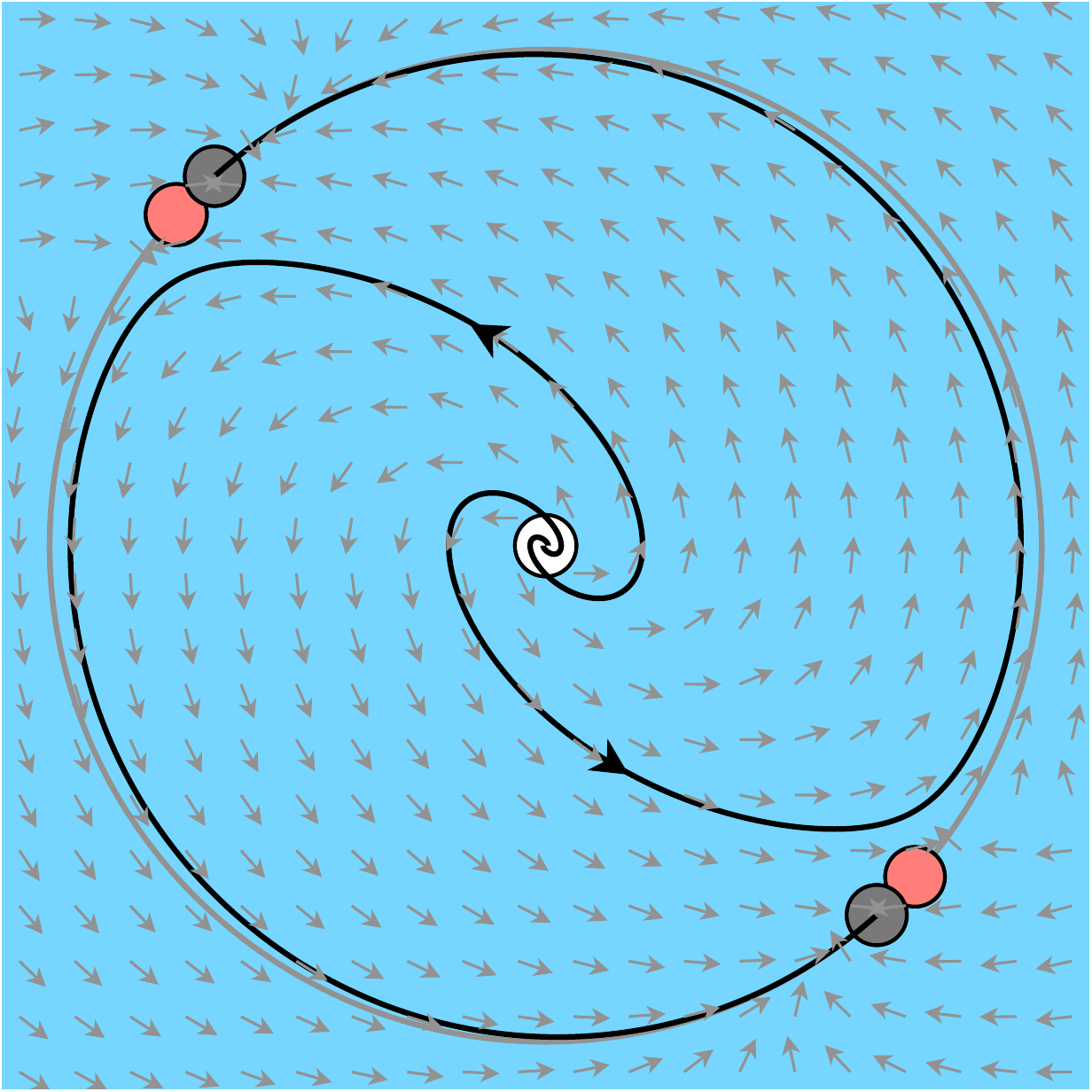} \\ \vspace{0.1cm}
    $b=0.400$
    \includegraphics[width=0.2\textwidth]{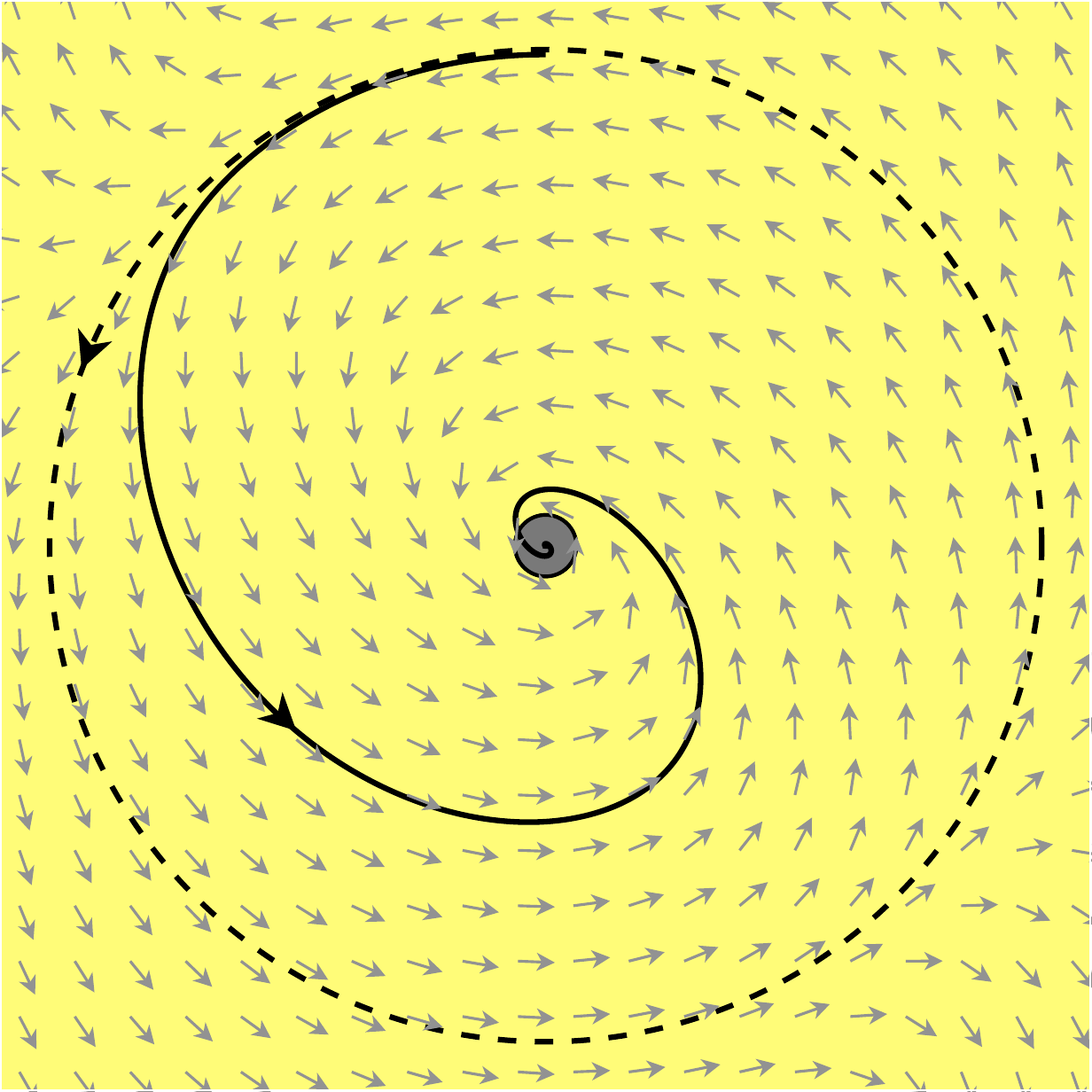}
    \includegraphics[width=0.2\textwidth]{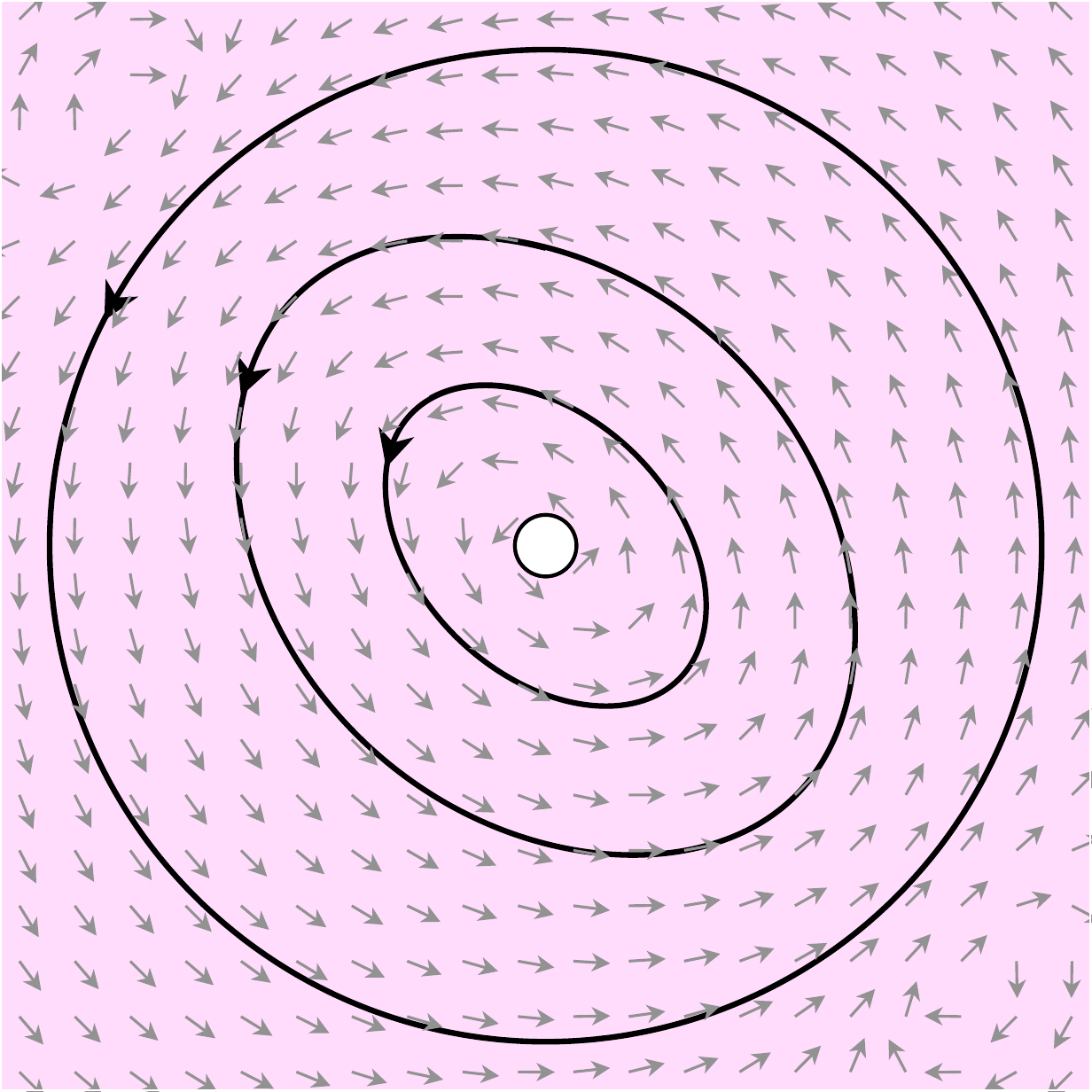}
    \includegraphics[width=0.2\textwidth]{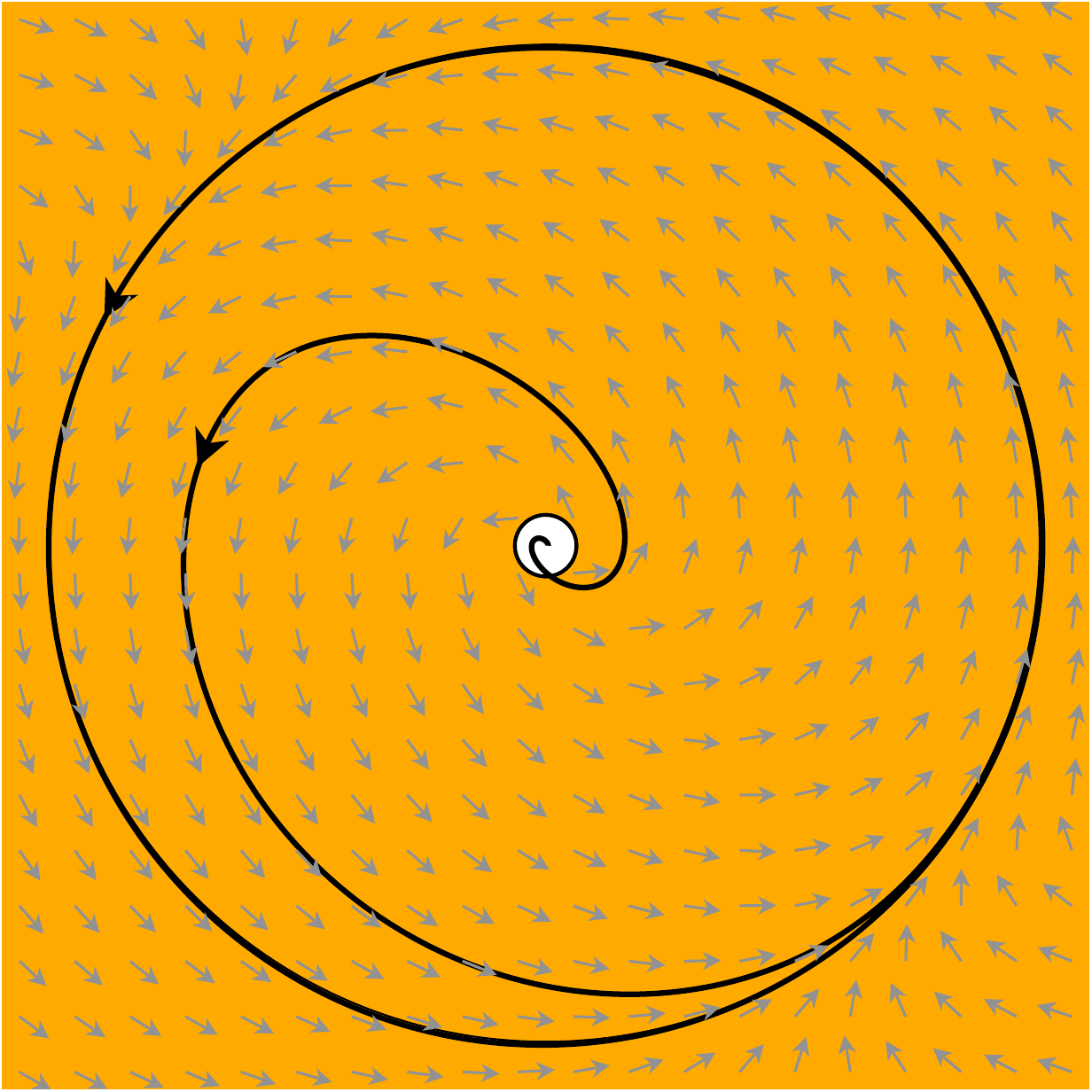}
    \caption{Phase portraits in the cartesian coordinates. Equilibrium points are represented by small circles in red (saddle), gray (sink), and white (source or center). In gray is the unitary circle $\rho = 1$. The background colors indicate the region in the parameter space according to Fig. \ref{fig:DBif} (right panel). 
    %From left to right $a = -0.5$, $a=0$, and $0.5$ are fixed while, from top to bottom $b = 1.01$, $b = 0.68$, $b = 0.503$, and $b = 0.4$ are fixed
    }
    \label{fig:1S}
\end{figure}

The condition $\dot{\rho}=0$ is fulfill with $\rho = 1$.
$\dot{\phi}=0$ is achieved when $2b\sin(2\phi) = -1$ or, i.e., $2b\cos(2\phi) = \pm\sqrt{4b^2-1}$.
Thence, the equilibria are $(1, \phi_3^+)$ and $(1, \phi_3^++\pi)$, for $\cos(2\phi) > 0$, and $(1, \phi_3^-)$ and $(1, \phi_3^-+\pi)$, for $\cos(2\phi) < 0$, where $\phi_3^+ = [-\pi/4, 0)$ and $\phi_3^- = (-\pi/2, -\pi/4]$ with Jacobian matrix (\ref{eq:3S}) 
\begin{equation}
J = 
\begin{pmatrix}
    -2a\pm \sqrt{4b^2-1} & \quad 0 \\
    -1 & \quad \pm2\sqrt{4b^2-1}
\end{pmatrix}_{(1, \phi^{\pm})} .
\label{eq:12S}
\end{equation}
The eigenvalues are 
\begin{eqnarray}
    \text{$\phi^+$:} \qquad \lambda_1^+ &=& -2a + \sqrt{4b^2-1} \label{eq:13S}\\
    \quad \lambda_2^+ &=& 2\sqrt{4b^2-1}\, , \label{eq:14S}
\end{eqnarray}
and
\begin{eqnarray} 
    \text{$\phi^-$:} \qquad \lambda_1^- &=& -2a - \sqrt{4b^2-1} \label{eq:15S}\\
     \quad \lambda_2^- &=& -2\sqrt{4b^2-1}\, . \label{eq:16S}
\end{eqnarray}
According the eigenvalues, for $b<1/2$ there is no equilibria at the unitary circle. 
The eigenvalues are complex and an oscillatory movement takes place with periodic angular velocity $\dot{\phi} = 1+2b\sin(2\phi)$. 
The stability of the equilibrium at the origin reflects the stability of the limit cycle. 
Thus, it is unstable if $a<0$ and asymptotically stable for $a>0$. 
For $a=0$, all eigenvalues are purely complex given rise to a degenerated {\it Andronov-Hopf} bifurcation, hence the limit cycle is Lyapunov stable (see Fig. \ref{fig:1S} with $b=0.4$).  

For $b=1/2$, $\lambda_2^\pm = 0$. 
The two pairs of equilibria emerge simultaneously by two {\it saddle-node heteroclinic} 
(SNH) bifurcations on the unitary circle: one with $\phi_3^+ = \phi_3^- = -\pi/4$ and the other out of phase by $\pi$.
Such bifurcations can also be seen as {\it saddle-node bifurcations on a limit cycle} (SNLC). 
Taking $b \gtrsim 0.5$, when $a<0$, $\lambda_{1,2}^+ >0$ and $\lambda_1^- \lambda_2^-<0$ what means that $(1,\phi_3^+)$ is a source node and $(1,\phi_3^-)$ is a saddle. 
The same is for the symmetric ones. 
Considering now $a>0$, $\lambda_{1,2}^- <0$ and $\lambda_1^+ \lambda_2^+<0$ what means that $(1,\phi^-)$ is a sink node and $(1,\phi^+)$ is a saddle. For these bifurcations, see Fig. \ref{fig:1S} whith $b = 0.503$). 

Each pair of equilibrium points separates along the unitary circle as $b$ grows. 
When $b \gtrsim \sqrt{a^2+1/4}$, if $a<0$, the saddle $(1, \phi^-)$ becomes a sink since $\lambda_1^-\lambda_1^- < 0$ while, if $a>0$, the saddle $(1, \phi^+)$ become a source as $\lambda_1^+\lambda_1^+ > 0$. 
Dynamically, the saddles $(1, \phi_3^\pm)$ collide with the "external" equilibria $(\rho>1,\phi_2^\pm)$ and exchange their stability.
This bifurcation happens at $b = \sqrt{a^2+1/4}$, where one of the eigenvalues $\lambda_1^\pm$ is null depending on the sign of $a$. 
This is the {\it transcritic} bifurcation [See Fig. \ref{fig:1S} with $a \neq 0$ and $b = 0.68$]. 
The special set $(a, b) = (0, 1/2)$ makes $\lambda_{1,2}^\pm = 0$ is another codimension two Bogdanov-Takens bifurcation that concentrates a saddle-node heteroclinic (SNH), 
transcritic, and a degenerate Andronov-Hopf bifurcations as shows Fig. \ref{fig:1S} with $a=0$ and $b=0.503$. 

%%%%%%%%%%%%%%%%%%%%%%%%%
%\input{Sections/6.1-Multiple/v-summary}

\subsection{Reduced model stability}

In Fig.\ref{fig:1S} we present an overview of the 
solutions for different regions in the parameter plane. 
In the left (center, right) column are presented phase portraits for 
$a=-0.5$ ($a=0, 0.5$) and several different values of $b$.
Equilibrium points are represented by small circles in
red (saddle), gray (sink), and white (source or center). 
In gray is the unitary circle $\rho$ = 1.
We notice that the study in Section \ref{Stablity}, concerning systems of two oscillators, aligns with those illustrated in Fig. \ref{fig:1S} for $b=0.503$ and $b=0.4$. 
Furthermore, it is observed that as $b$ increases beyond the SNH (or SNLC) bifurcation point, the skin and saddle (for $a>0$) and the source and saddle (for $a<0$) migrate in opposing directions relative to the positions $3\pi/4$ and $-\pi/4$. 
This behavior suggests that the observed pitchfork and transcritical bifurcations stem from collective interactions within the system. 

In some portraits, we can observe some attractors outside
the circle $\rho=1$. Those points are solutions of
Eq.(\ref{eqZ2}) but unreachable for the case of the Kuramoto model, since for any configuration of phase oscillators
$\rho \leq 1$.

\begin{figure}[htb]
    \centering
    \includegraphics[width=0.45\textwidth]{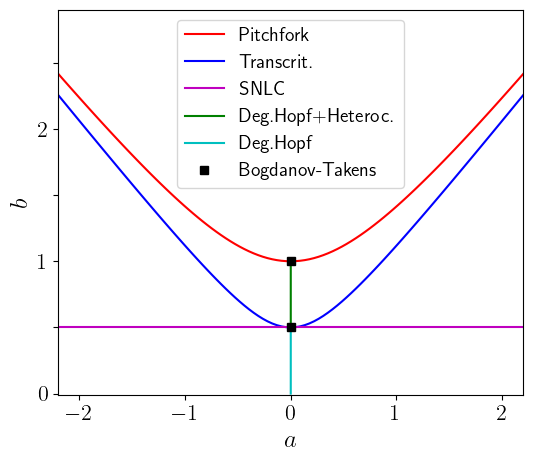}    %%DB_Kuramoto_QSR.pdf}
    \includegraphics[width=0.45\textwidth]{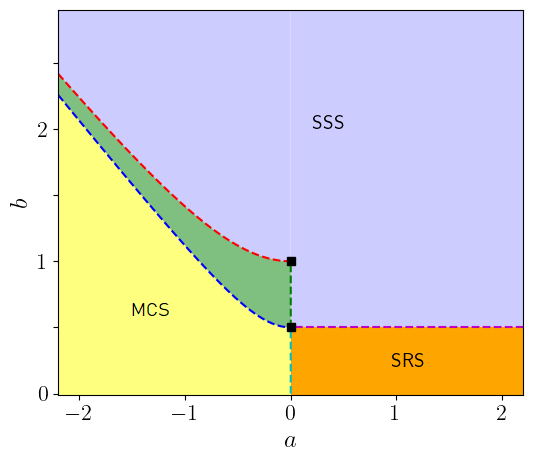}     %%PhaseDiag_full.pdf}
    \caption{Left: Bifurcations in the parameter plane. Right: phase diagram with 
    the collective states observed in each region of the parameter plane.
    Regions of the phase diagram are delimited by parts of the bifurcation curves.}
    \label{fig:DBif}
\end{figure}

In the left panel of Fig.\ref{fig:DBif} we summarize all these bifurcation curves/points in the parameter plane. 
In the right panel we present the phase diagram with the collective states observed in each region of the parameter 
plane, such states are detailed in the next section.

%%%%%%%%%%%%%%%%%%%%%%%%%%%%%%%%%%%%%%%%%%%%%%%%%%%%%%%%%%%%%%%%%%%%%%%%%%%%%%%%%%%%%%%%%%%%%%%%%%%%%%%%%%%%%%%%%%%%%%%%%%%%%
%\input{Sections/7-Collective}

\section{Collective Dynamics and Phase Diagram}
\label{sec5}

We also performed numerical integrations of the equations (\ref{eq:kqsr})
with identical oscillators in order
analyze the collective dynamics of the network.
By exploring numerically the parameter space (Fig.\ref{fig:DBif}) we identified 
three main types of collective states for different regions of the 
bifurcation/phase diagram:  synchronized stationary, 
synchronized rotational and multi-cluster states, which are detailed below.

{\bf Synchronized rotational state} (SRS) In the parameter plane:
    for $a>0$, below the purple line, i.e.\ for $b<1/2$, corresponding to the orange 
    region in the right panel of Fig.\ref{fig:DBif}. The last phase portrait
    (with orange background) in Fig.\ref{fig:1S} shows that the origin ($\rho=0$)
    is a repeller and the circle with $\rho=1$ is a stable limit cycle.
    The collective dynamics will manifest with all oscillators in phase
    rotating in the laboratory frame according to Eq.(\ref{eq:2S}). This state
    resambles the "Standing wave state" (SWS) of Ref.\cite{ChandrasekarPRE2020}
    and the "Oscillatory synchronized" (OS) state of Ref.\cite{Manoranjani2021}, 
    but in both papers the order parameter $R(t)$ 
    oscillates with time and its time average is in general 
    $0 < \langle R\rangle \lesssim 1$. 
    Here the SRS is characterized by $R(t) = 1$ achieved after some initial transient.
\footnote{Let us remember that $\rho$ and $R$ are distinct objects. The variable $\rho$
can be seen as the prediction of the reduced model (obtained through the WS formalism) 
for the amplitude $R$ of the Kuramoto order parameter, which can be measured
directly from numerical integration of the $N$ Kuramoto equations (\ref{eq:kqsr}). 
} 

{\bf Synchronized stationary state} (SSS) In the bifurcation/phase diagram 
(the green and blue regions in the right panel of 
    Fig.\ref{fig:DBif}):
    (i) for $a<0$, above the blue curve (transcritical bifurcations);
    (ii) for $a>0$, above the purple line (SNLC bifurcations), i.e.\ for
    $b>1/2$.
    The network has all oscillators in phase, $R(t)=1$, pointing in a fixed direction 
    of the laboratory frame, with angle given by one of the (two) stable solutions 
    of equation $1+2b\sin(2\phi^{*})=0$. Our SSS is similar to the (homonym) 
    SSS observed in Ref.\cite{ChandrasekarPRE2020}
    and the "Oscillation death" (OD) state of Ref.\cite{Manoranjani2021}, 
    but in both papers the time average of order parameter 
    is in general $0 < \langle R\rangle \lesssim 1$.
In the green region (right panel of Fig.\ref{fig:DBif}) multi-cluster and synchronized 
stationary states coexist. The first phase portrait of the left column (with green background) 
in Fig.\ref{fig:1S} makes explicit the coexistence of three attractors: the origin and two 
diametrically opposite points on the circle $\rho=1$, and their basins of attraction are 
defined by invariant manifolds of the pair of saddles.

{\bf Multi-cluster state} (MCS) Such state correspond to the solution
    $\rho=0$. In the parameter plane it is the region for $a<0$ and below the (red) pitchfork 
    bifurcation curve, i.e.\ the yellow and green regions in the right panel of 
    Fig.\ref{fig:DBif}. Here we consider a cluster as a group with two or more nodes 
    rotating in phase.
    In general a state composed of two or more clusters with contiguous nodes in each 
    cluster is stable and for a random initial configuration of the network nodes, 
    the dynamics tends to form clusters with non-contiguous nodes in each cluster. 

    Multi-cluster states are not reported in Refs.\cite{ChandrasekarPRE2020, Manoranjani2021},
    but our MCS are roughly similar to the rotating clusters observed recently in networks of Kuramoto oscillators 
with higher-order (triadic) interactions and phase-lag \cite{Jalan_PRE2024}, but there the oscillators 
are nonidentical and as a result oscillators inside a cluster are phase-locked but not in phase and 
there are also some oscillators drifting, i.e. do not belong to none of the clusters, as shown in Fig.~1(a) 
of Ref.\cite{Jalan_PRE2024}. In contrast, in our multi-cluster state there is no oscillator drifting 
and in each cluster the oscillators are in phase. On the other hand, the multi-cluster states we observed
in this model are very similar to the (frequency) clusters that appear in the neuron model studied
in Ref.\cite{Rohr2019}.

\begin{figure}[ht]
    \centering
    \includegraphics[width=0.65\linewidth]{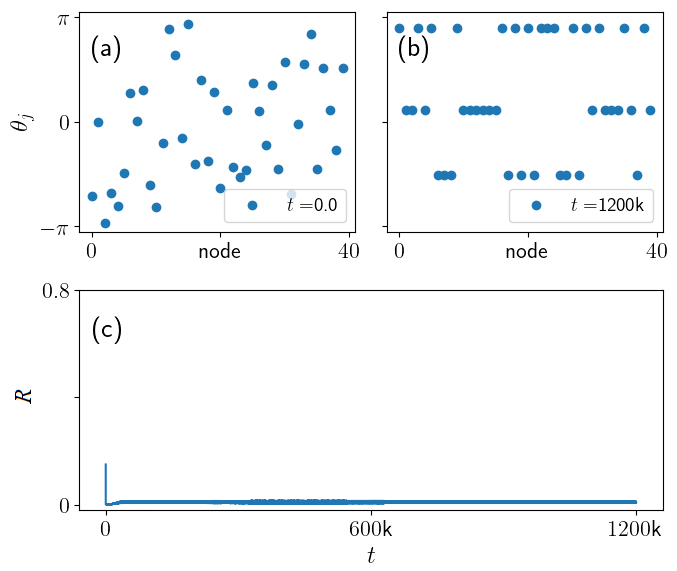}
    \caption{Starting with (a) random initial state, the network (with 40 nodes) converges 
    to some (b) multi-cluster state. (c) Behavior of the Kuramoto order parameter $R(t)$ . 
    (Parameters: $a = -0.5$ and $b = 0.8$)} 
    \label{fig:Random-MC}
\end{figure}

In Fig.\ref{fig:Random-MC} we present the results of numerical integration of Eqs.~(\ref{eq:kqsr})
for $N=40$ nodes; $\omega_j=1,\,\forall\,j$; parameters $a=-0.5,\, b=0.8$ and (a) random
state as initial condition. The (b) final state has three clusters with non-contiguous 
nodes in each cluster. In Fig.\ref{fig:Random-MC}(c) we present the behavior of the 
(absolute value of) Kuramoto order parameter $R$ as a function of time.

\subsection{``Virtual'' attractors}

The WS theory reduced the dimension of the original model 
Eq.(\ref{eq:kqsr}) and provided us with a new set of equations
(\ref{eqXY}) which in turn have solutions outside the circle of radius 
$\rho = 1$. However the collective behavior of the Kuramoto
oscillators constrains $\rho$ so that it is less than or 
equal to 1. On the other hand it is possible observe the
``virtual'' attraction of the outsiders (solutions with 
$\rho\gtrsim 1$) by comparing the transient time for
Eq.(\ref{eq:kqsr}) with different values of some parameter. 

\begin{figure}[ht]
    \centering
    \includegraphics[width=0.65\linewidth]{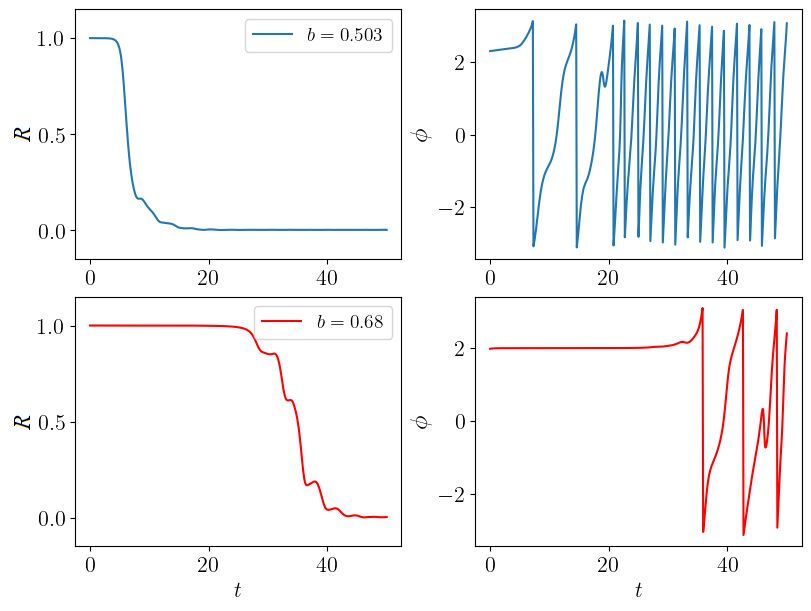}
    \caption{Results of simulations of Eq.(\ref{eq:kqsr}) for $a=-0.5$ and $b=0.503, \, 0.68$.
    In both cases the initial conditions are close to the saddles, but for $b=0.68$ the transient
    time is greater, supposedly because of the existence of an outside (with $\rho \gtrsim 1$) attractor 
    very close to the saddle. Notice the second and third phase portraits in the left column of Fig.\ref{fig:1S}.}
    \label{fig:virtual}
\end{figure}

Let us consider the second ($b=0.68$) and third ($b=0.503$)
phase portraits in the left column ($a=-0.5$) of
Fig.\ref{fig:1S}. For $b=0.68$, Eqs.(\ref{eqXY}) have a pair of
stable solutions with $\rho\gtrsim 1$ and they are very
close to the pair of saddles on the circle with $\rho=1$.
On the other hand for $b=0.503$, Eqs.(\ref{eqXY}) there
are no attractors nearby the saddles. We present in 
Fig.\ref{fig:virtual} the results of simulations of
Eq.(\ref{eq:kqsr}) for $a=-0.5$ and $b=0.503, \, 0.68$. 
For both cases we measured the absolute value and the 
angle of the Kuramoto order parameter $Z = R e^{i\varphi}$,
with the following initial condition:
$\theta_j (0) = \phi^* + \delta_j,\, \forall\,j$,
where $\phi^*$ is the (saddle) solution of 
$1 + 2 b \sin(2\phi^*) = 0$ and $\delta_j$ is a
random noise between -0.005 and 0.005. 
We can observe that for $b=0.68$ the presence of
an attractor, with $\rho \approx R \gtrsim 1$ and very close to the saddle,
seems to increase the initial transient time nearby the saddle.

%%%%%%%%%%%%%%%%%%%%%%%%%
%\input{Sections/8-Remarks}

\section{Final remarks}
\label{sec6}

In this work we have analyzed the effects of interactions that break rotational symmetry on 
a network of Kuramoto oscillators.  The model we analysed here can also be seen as a 
generalization of the model proposed by \cite{Ariaratnam2001} to describe populations 
of pulse-coupled biological oscillators. 
Initially, we study the effect of the symmetry breaking term in the Kuramoto model for two oscillators.
Next, since we considered the case of a population of 
identical oscillators, we used the Watanabe-Strogatz (WS) approach in order to reduce the 
dimensionality of the original model (\ref{eq:kqsr}). Then we analysed the ODEs for the 
complex variable $z$ (approximately equal to the Kuramoto order parameter), provided by the 
WS framework, obtaining their solutions and respective stabilities and a rich landscape of 
bifurcations in the parameter space.

We also performed extensive numerical studies of (\ref{eq:kqsr}) in order to compare the 
collective behavior of oscillators with the results (solutions) for 
$z(t)=\rho(t).e^{i \phi(t)}$ , resulting in the phase diagram of Fig.\ref{fig:DBif} (right panel). 
We observed three main types of collective behavior. Two of them (SSS and SRS) are roughly similar 
to the collective states observed in Refs.\cite{ChandrasekarPRE2020, Manoranjani2021}, but 
multi-cluster states MCS are not reported in none of them. 
Recently rotating clusters were observed in networks of Kuramoto oscillators 
with higher-order (triadic) interactions (and phase-lag) \cite{Jalan_PRE2024}
but such clusters have some differences compared to the MCS described above.
On the other hand our MCS behave very similarly to the (frequency) clusters observed 
in the model with Hodgkin-Huxley neurons of Ref.\cite{Rohr2019}.

%%%%%%%%%%%%%%%%%%%%%%%%%
%\input{Sections/9-Acknowledgments}

\section*{Acknowledgments}

The authors are grateful for fruitful discussions with Everton S. Medeiros.
AM and ROMT acknowledge the financial support by the São Paulo Research
Foundation (process FAPESP 2023/08144-3 and 2024/06718-5) and by the 
National Council for Scientific and Technological Development (process CNPq 408522/2023-2).

%%%%%%%%%%%%%%%%%%%%%%%%%%%%%%%%%%
\bibliography{refs.bib}

%apsrev4-2.bst 2019-01-14 (MD) hand-edited version of apsrev4-1.bst
%Control: key (0)
%Control: author (72) initials jnrlst
%Control: editor formatted (1) identically to author
%Control: production of article title (-1) disabled
%Control: page (0) single
%Control: year (1) truncated
%Control: production of eprint (0) enabled
\begin{thebibliography}{29}%
\makeatletter
\providecommand \@ifxundefined [1]{%
 \@ifx{#1\undefined}
}%
\providecommand \@ifnum [1]{%
 \ifnum #1\expandafter \@firstoftwo
 \else \expandafter \@secondoftwo
 \fi
}%
\providecommand \@ifx [1]{%
 \ifx #1\expandafter \@firstoftwo
 \else \expandafter \@secondoftwo
 \fi
}%
\providecommand \natexlab [1]{#1}%
\providecommand \enquote  [1]{``#1''}%
\providecommand \bibnamefont  [1]{#1}%
\providecommand \bibfnamefont [1]{#1}%
\providecommand \citenamefont [1]{#1}%
\providecommand \href@noop [0]{\@secondoftwo}%
\providecommand \href [0]{\begingroup \@sanitize@url \@href}%
\providecommand \@href[1]{\@@startlink{#1}\@@href}%
\providecommand \@@href[1]{\endgroup#1\@@endlink}%
\providecommand \@sanitize@url [0]{\catcode `\\12\catcode `\$12\catcode `\&12\catcode `\#12\catcode `\^12\catcode `\_12\catcode `\%12\relax}%
\providecommand \@@startlink[1]{}%
\providecommand \@@endlink[0]{}%
\providecommand \url  [0]{\begingroup\@sanitize@url \@url }%
\providecommand \@url [1]{\endgroup\@href {#1}{\urlprefix }}%
\providecommand \urlprefix  [0]{URL }%
\providecommand \Eprint [0]{\href }%
\providecommand \doibase [0]{https://doi.org/}%
\providecommand \selectlanguage [0]{\@gobble}%
\providecommand \bibinfo  [0]{\@secondoftwo}%
\providecommand \bibfield  [0]{\@secondoftwo}%
\providecommand \translation [1]{[#1]}%
\providecommand \BibitemOpen [0]{}%
\providecommand \bibitemStop [0]{}%
\providecommand \bibitemNoStop [0]{.\EOS\space}%
\providecommand \EOS [0]{\spacefactor3000\relax}%
\providecommand \BibitemShut  [1]{\csname bibitem#1\endcsname}%
\let\auto@bib@innerbib\@empty
%</preamble>
\bibitem [{\citenamefont {Winfree}(1967)}]{Winfree1967}%
  \BibitemOpen
  \bibfield  {author} {\bibinfo {author} {\bibfnamefont {A.~T.}\ \bibnamefont {Winfree}},\ }\href@noop {} {\bibfield  {journal} {\bibinfo  {journal} {J. Theor. Biol.}\ }\textbf {\bibinfo {volume} {16}},\ \bibinfo {pages} {15} (\bibinfo {year} {1967})}\BibitemShut {NoStop}%
\bibitem [{\citenamefont {Kuramoto}(1975)}]{Kuramoto1975}%
  \BibitemOpen
  \bibfield  {author} {\bibinfo {author} {\bibfnamefont {Y.}~\bibnamefont {Kuramoto}},\ }in\ \href@noop {} {\emph {\bibinfo {booktitle} {International symposium on mathematical problems in theoretical physics}}}\ (\bibinfo {organization} {Springer},\ \bibinfo {year} {1975})\ pp.\ \bibinfo {pages} {420--422}\BibitemShut {NoStop}%
\bibitem [{\citenamefont {Kuramoto}(1984)}]{Kuramoto1984}%
  \BibitemOpen
  \bibfield  {author} {\bibinfo {author} {\bibfnamefont {Y.}~\bibnamefont {Kuramoto}},\ }\href@noop {} {\emph {\bibinfo {title} {Chemical Oscillations, Waves, and Turbulence}}}\ (\bibinfo  {publisher} {Springer-Verlag},\ \bibinfo {address} {Berlin},\ \bibinfo {year} {1984})\BibitemShut {NoStop}%
\bibitem [{\citenamefont {Mihara}\ and\ \citenamefont {Medrano-T}(2019)}]{Mihara2019}%
  \BibitemOpen
  \bibfield  {author} {\bibinfo {author} {\bibfnamefont {A.}~\bibnamefont {Mihara}}\ and\ \bibinfo {author} {\bibfnamefont {R.~O.}\ \bibnamefont {Medrano-T}},\ }\href@noop {} {\bibfield  {journal} {\bibinfo  {journal} {Nonlinear Dyn.}\ }\textbf {\bibinfo {volume} {98}},\ \bibinfo {pages} {539} (\bibinfo {year} {2019})}\BibitemShut {NoStop}%
\bibitem [{\citenamefont {Budzinski}\ \emph {et~al.}(2022)\citenamefont {Budzinski}, \citenamefont {Nguyen}, \citenamefont {{\DJ}o{\`a}n}, \citenamefont {Min{\'a}{\v{c}}}, \citenamefont {Sejnowski},\ and\ \citenamefont {Muller}}]{budzinski2022geometry}%
  \BibitemOpen
  \bibfield  {author} {\bibinfo {author} {\bibfnamefont {R.~C.}\ \bibnamefont {Budzinski}}, \bibinfo {author} {\bibfnamefont {T.~T.}\ \bibnamefont {Nguyen}}, \bibinfo {author} {\bibfnamefont {J.}~\bibnamefont {{\DJ}o{\`a}n}}, \bibinfo {author} {\bibfnamefont {J.}~\bibnamefont {Min{\'a}{\v{c}}}}, \bibinfo {author} {\bibfnamefont {T.~J.}\ \bibnamefont {Sejnowski}},\ and\ \bibinfo {author} {\bibfnamefont {L.~E.}\ \bibnamefont {Muller}},\ }\href@noop {} {\bibfield  {journal} {\bibinfo  {journal} {Chaos}\ }\textbf {\bibinfo {volume} {32}} (\bibinfo {year} {2022})}\BibitemShut {NoStop}%
\bibitem [{\citenamefont {Budzinski}\ \emph {et~al.}(2023)\citenamefont {Budzinski}, \citenamefont {Nguyen}, \citenamefont {Benigno}, \citenamefont {{\DJ}o{\`a}n}, \citenamefont {Min{\'a}{\v{c}}}, \citenamefont {Sejnowski},\ and\ \citenamefont {Muller}}]{budzinski2023analytical}%
  \BibitemOpen
  \bibfield  {author} {\bibinfo {author} {\bibfnamefont {R.~C.}\ \bibnamefont {Budzinski}}, \bibinfo {author} {\bibfnamefont {T.~T.}\ \bibnamefont {Nguyen}}, \bibinfo {author} {\bibfnamefont {G.~B.}\ \bibnamefont {Benigno}}, \bibinfo {author} {\bibfnamefont {J.}~\bibnamefont {{\DJ}o{\`a}n}}, \bibinfo {author} {\bibfnamefont {J.}~\bibnamefont {Min{\'a}{\v{c}}}}, \bibinfo {author} {\bibfnamefont {T.~J.}\ \bibnamefont {Sejnowski}},\ and\ \bibinfo {author} {\bibfnamefont {L.~E.}\ \bibnamefont {Muller}},\ }\href@noop {} {\bibfield  {journal} {\bibinfo  {journal} {Phys. Rev. Res.}\ }\textbf {\bibinfo {volume} {5}},\ \bibinfo {pages} {013159} (\bibinfo {year} {2023})}\BibitemShut {NoStop}%
\bibitem [{\citenamefont {Mihara}\ \emph {et~al.}(2025)\citenamefont {Mihara}, \citenamefont {Kuwana}, \citenamefont {Budzinski}, \citenamefont {Muller},\ and\ \citenamefont {Medrano-T}}]{Mihara2025}%
  \BibitemOpen
  \bibfield  {author} {\bibinfo {author} {\bibfnamefont {A.}~\bibnamefont {Mihara}}, \bibinfo {author} {\bibfnamefont {C.~M.}\ \bibnamefont {Kuwana}}, \bibinfo {author} {\bibfnamefont {R.~C.}\ \bibnamefont {Budzinski}}, \bibinfo {author} {\bibfnamefont {L.~E.}\ \bibnamefont {Muller}},\ and\ \bibinfo {author} {\bibfnamefont {R.~O.}\ \bibnamefont {Medrano-T}},\ }\href@noop {} {\bibfield  {journal} {\bibinfo  {journal} {Chaos}\ }\textbf {\bibinfo {volume} {35}} (\bibinfo {year} {2025})}\BibitemShut {NoStop}%
\bibitem [{\citenamefont {Sinha}\ \emph {et~al.}(2025)\citenamefont {Sinha}, \citenamefont {Jain}, \citenamefont {Mihara}, \citenamefont {Medrano-T}, \citenamefont {Min{\'a}{\v{c}}}, \citenamefont {Muller},\ and\ \citenamefont {Budzinski}}]{Sinha2025}%
  \BibitemOpen
  \bibfield  {author} {\bibinfo {author} {\bibfnamefont {Y.}~\bibnamefont {Sinha}}, \bibinfo {author} {\bibfnamefont {P.~B.}\ \bibnamefont {Jain}}, \bibinfo {author} {\bibfnamefont {A.}~\bibnamefont {Mihara}}, \bibinfo {author} {\bibfnamefont {R.~O.}\ \bibnamefont {Medrano-T}}, \bibinfo {author} {\bibfnamefont {J.}~\bibnamefont {Min{\'a}{\v{c}}}}, \bibinfo {author} {\bibfnamefont {L.~E.}\ \bibnamefont {Muller}},\ and\ \bibinfo {author} {\bibfnamefont {R.~C.}\ \bibnamefont {Budzinski}},\ }\href@noop {} {\bibfield  {journal} {\bibinfo  {journal} {arXiv preprint arXiv:2504.06377}\ } (\bibinfo {year} {2025})}\BibitemShut {NoStop}%
\bibitem [{\citenamefont {Wiley}\ \emph {et~al.}(2006)\citenamefont {Wiley}, \citenamefont {Strogatz},\ and\ \citenamefont {Girvan}}]{Wiley2006}%
  \BibitemOpen
  \bibfield  {author} {\bibinfo {author} {\bibfnamefont {D.~A.}\ \bibnamefont {Wiley}}, \bibinfo {author} {\bibfnamefont {S.~H.}\ \bibnamefont {Strogatz}},\ and\ \bibinfo {author} {\bibfnamefont {M.}~\bibnamefont {Girvan}},\ }\href@noop {} {\bibfield  {journal} {\bibinfo  {journal} {Chaos}\ }\textbf {\bibinfo {volume} {16}},\ \bibinfo {pages} {015103} (\bibinfo {year} {2006})}\BibitemShut {NoStop}%
\bibitem [{\citenamefont {Delabays}\ \emph {et~al.}(2017)\citenamefont {Delabays}, \citenamefont {Tyloo},\ and\ \citenamefont {Jacquod}}]{Delabays2017}%
  \BibitemOpen
  \bibfield  {author} {\bibinfo {author} {\bibfnamefont {R.}~\bibnamefont {Delabays}}, \bibinfo {author} {\bibfnamefont {M.}~\bibnamefont {Tyloo}},\ and\ \bibinfo {author} {\bibfnamefont {P.}~\bibnamefont {Jacquod}},\ }\href@noop {} {\bibfield  {journal} {\bibinfo  {journal} {Chaos}\ }\textbf {\bibinfo {volume} {27}},\ \bibinfo {pages} {103109} (\bibinfo {year} {2017})}\BibitemShut {NoStop}%
\bibitem [{\citenamefont {Mihara}\ \emph {et~al.}(2022{\natexlab{a}})\citenamefont {Mihara}, \citenamefont {Zaks}, \citenamefont {Macau},\ and\ \citenamefont {Medrano-T}}]{Mihara2022}%
  \BibitemOpen
  \bibfield  {author} {\bibinfo {author} {\bibfnamefont {A.}~\bibnamefont {Mihara}}, \bibinfo {author} {\bibfnamefont {M.}~\bibnamefont {Zaks}}, \bibinfo {author} {\bibfnamefont {E.~E.~N.}\ \bibnamefont {Macau}},\ and\ \bibinfo {author} {\bibfnamefont {R.~O.}\ \bibnamefont {Medrano-T}},\ }\href@noop {} {\bibfield  {journal} {\bibinfo  {journal} {Phys. Rev. E}\ }\textbf {\bibinfo {volume} {105}},\ \bibinfo {pages} {L052202} (\bibinfo {year} {2022}{\natexlab{a}})}\BibitemShut {NoStop}%
\bibitem [{\citenamefont {Townsend}\ \emph {et~al.}(2020)\citenamefont {Townsend}, \citenamefont {Stillman},\ and\ \citenamefont {Strogatz}}]{townsend2020dense}%
  \BibitemOpen
  \bibfield  {author} {\bibinfo {author} {\bibfnamefont {A.}~\bibnamefont {Townsend}}, \bibinfo {author} {\bibfnamefont {M.}~\bibnamefont {Stillman}},\ and\ \bibinfo {author} {\bibfnamefont {S.~H.}\ \bibnamefont {Strogatz}},\ }\href@noop {} {\bibfield  {journal} {\bibinfo  {journal} {Chaos}\ }\textbf {\bibinfo {volume} {30}} (\bibinfo {year} {2020})}\BibitemShut {NoStop}%
\bibitem [{\citenamefont {Mihara}\ \emph {et~al.}(2022{\natexlab{b}})\citenamefont {Mihara}, \citenamefont {Medeiros}, \citenamefont {Zakharova},\ and\ \citenamefont {Medrano-T}}]{mihara2022sparsity}%
  \BibitemOpen
  \bibfield  {author} {\bibinfo {author} {\bibfnamefont {A.}~\bibnamefont {Mihara}}, \bibinfo {author} {\bibfnamefont {E.~S.}\ \bibnamefont {Medeiros}}, \bibinfo {author} {\bibfnamefont {A.}~\bibnamefont {Zakharova}},\ and\ \bibinfo {author} {\bibfnamefont {R.~O.}\ \bibnamefont {Medrano-T}},\ }\href@noop {} {\bibfield  {journal} {\bibinfo  {journal} {Chaos}\ }\textbf {\bibinfo {volume} {32}} (\bibinfo {year} {2022}{\natexlab{b}})}\BibitemShut {NoStop}%
\bibitem [{\citenamefont {Acebr\'on}\ \emph {et~al.}(2005)\citenamefont {Acebr\'on}, \citenamefont {Bonilla}, \citenamefont {Vicente}, \citenamefont {Ritort},\ and\ \citenamefont {Spigler}}]{Acebron2005}%
  \BibitemOpen
  \bibfield  {author} {\bibinfo {author} {\bibfnamefont {J.~A.}\ \bibnamefont {Acebr\'on}}, \bibinfo {author} {\bibfnamefont {L.~L.}\ \bibnamefont {Bonilla}}, \bibinfo {author} {\bibfnamefont {C.~J.~P.}\ \bibnamefont {Vicente}}, \bibinfo {author} {\bibfnamefont {F.}~\bibnamefont {Ritort}},\ and\ \bibinfo {author} {\bibfnamefont {R.}~\bibnamefont {Spigler}},\ }\href@noop {} {\bibfield  {journal} {\bibinfo  {journal} {Rev. Mod. Phys.}\ }\textbf {\bibinfo {volume} {77}},\ \bibinfo {pages} {137} (\bibinfo {year} {2005})}\BibitemShut {NoStop}%
\bibitem [{\citenamefont {Rodrigues}\ \emph {et~al.}(2016)\citenamefont {Rodrigues}, \citenamefont {Peron}, \citenamefont {Ji},\ and\ \citenamefont {Kurths}}]{rodrigues2016kuramoto}%
  \BibitemOpen
  \bibfield  {author} {\bibinfo {author} {\bibfnamefont {F.~A.}\ \bibnamefont {Rodrigues}}, \bibinfo {author} {\bibfnamefont {T.~K.~D.}\ \bibnamefont {Peron}}, \bibinfo {author} {\bibfnamefont {P.}~\bibnamefont {Ji}},\ and\ \bibinfo {author} {\bibfnamefont {J.}~\bibnamefont {Kurths}},\ }\href@noop {} {\bibfield  {journal} {\bibinfo  {journal} {Phys. Rep.}\ }\textbf {\bibinfo {volume} {610}},\ \bibinfo {pages} {1} (\bibinfo {year} {2016})}\BibitemShut {NoStop}%
\bibitem [{\citenamefont {Antonsen{ }Jr}\ \emph {et~al.}(2008)\citenamefont {Antonsen{ }Jr}, \citenamefont {Faghih}, \citenamefont {Girvan}, \citenamefont {Ott},\ and\ \citenamefont {Platig}}]{Antonsen2008}%
  \BibitemOpen
  \bibfield  {author} {\bibinfo {author} {\bibfnamefont {T.~M.}\ \bibnamefont {Antonsen{ }Jr}}, \bibinfo {author} {\bibfnamefont {R.~T.}\ \bibnamefont {Faghih}}, \bibinfo {author} {\bibfnamefont {M.}~\bibnamefont {Girvan}}, \bibinfo {author} {\bibfnamefont {E.}~\bibnamefont {Ott}},\ and\ \bibinfo {author} {\bibfnamefont {J.}~\bibnamefont {Platig}},\ }\href@noop {} {\bibfield  {journal} {\bibinfo  {journal} {Chaos}\ }\textbf {\bibinfo {volume} {18}},\ \bibinfo {pages} {037112} (\bibinfo {year} {2008})}\BibitemShut {NoStop}%
\bibitem [{\citenamefont {O’Keeffe}\ \emph {et~al.}(2017)\citenamefont {O’Keeffe}, \citenamefont {Hong},\ and\ \citenamefont {Strogatz}}]{Keeffe2017}%
  \BibitemOpen
  \bibfield  {author} {\bibinfo {author} {\bibfnamefont {K.~P.}\ \bibnamefont {O’Keeffe}}, \bibinfo {author} {\bibfnamefont {H.}~\bibnamefont {Hong}},\ and\ \bibinfo {author} {\bibfnamefont {S.~H.}\ \bibnamefont {Strogatz}},\ }\href@noop {} {\bibfield  {journal} {\bibinfo  {journal} {Nat. Commun.}\ }\textbf {\bibinfo {volume} {8}},\ \bibinfo {pages} {1504} (\bibinfo {year} {2017})}\BibitemShut {NoStop}%
\bibitem [{\citenamefont {Osaka}(2017)}]{Osaka2017}%
  \BibitemOpen
  \bibfield  {author} {\bibinfo {author} {\bibfnamefont {M.}~\bibnamefont {Osaka}},\ }\href@noop {} {\bibfield  {journal} {\bibinfo  {journal} {Appl. Math.}\ }\textbf {\bibinfo {volume} {8}},\ \bibinfo {pages} {1227} (\bibinfo {year} {2017})}\BibitemShut {NoStop}%
\bibitem [{\citenamefont {Wiesenfeld}\ \emph {et~al.}(1998)\citenamefont {Wiesenfeld}, \citenamefont {Colet},\ and\ \citenamefont {Strogatz}}]{Wiesenfeld1998}%
  \BibitemOpen
  \bibfield  {author} {\bibinfo {author} {\bibfnamefont {K.}~\bibnamefont {Wiesenfeld}}, \bibinfo {author} {\bibfnamefont {P.}~\bibnamefont {Colet}},\ and\ \bibinfo {author} {\bibfnamefont {S.~H.}\ \bibnamefont {Strogatz}},\ }\href@noop {} {\bibfield  {journal} {\bibinfo  {journal} {Phys. Rev. E}\ }\textbf {\bibinfo {volume} {57}},\ \bibinfo {pages} {1563} (\bibinfo {year} {1998})}\BibitemShut {NoStop}%
\bibitem [{\citenamefont {D{\"o}rfler}\ \emph {et~al.}(2013)\citenamefont {D{\"o}rfler}, \citenamefont {Chertkov},\ and\ \citenamefont {Bullo}}]{Dorfler2013}%
  \BibitemOpen
  \bibfield  {author} {\bibinfo {author} {\bibfnamefont {F.}~\bibnamefont {D{\"o}rfler}}, \bibinfo {author} {\bibfnamefont {M.}~\bibnamefont {Chertkov}},\ and\ \bibinfo {author} {\bibfnamefont {F.}~\bibnamefont {Bullo}},\ }\href@noop {} {\bibfield  {journal} {\bibinfo  {journal} {Proc. Natl. Acad. Sci. U.S.A.}\ }\textbf {\bibinfo {volume} {110}},\ \bibinfo {pages} {2005} (\bibinfo {year} {2013})}\BibitemShut {NoStop}%
\bibitem [{\citenamefont {Vasudevan}\ \emph {et~al.}(2015)\citenamefont {Vasudevan}, \citenamefont {Cavers},\ and\ \citenamefont {Ware}}]{Vasudevan2015}%
  \BibitemOpen
  \bibfield  {author} {\bibinfo {author} {\bibfnamefont {K.}~\bibnamefont {Vasudevan}}, \bibinfo {author} {\bibfnamefont {M.}~\bibnamefont {Cavers}},\ and\ \bibinfo {author} {\bibfnamefont {A.}~\bibnamefont {Ware}},\ }\href {https://doi.org/10.5194/npg-22-499-2015} {\bibfield  {journal} {\bibinfo  {journal} {Nonlin. Processes Geophys.}\ }\textbf {\bibinfo {volume} {22}},\ \bibinfo {pages} {499} (\bibinfo {year} {2015})}\BibitemShut {NoStop}%
\bibitem [{\citenamefont {Chandrasekar}\ \emph {et~al.}(2020)\citenamefont {Chandrasekar}, \citenamefont {Manoranjani},\ and\ \citenamefont {Gupta}}]{ChandrasekarPRE2020}%
  \BibitemOpen
  \bibfield  {author} {\bibinfo {author} {\bibfnamefont {V.~K.}\ \bibnamefont {Chandrasekar}}, \bibinfo {author} {\bibfnamefont {M.}~\bibnamefont {Manoranjani}},\ and\ \bibinfo {author} {\bibfnamefont {S.}~\bibnamefont {Gupta}},\ }\href {https://doi.org/10.1103/PhysRevE.102.012206} {\bibfield  {journal} {\bibinfo  {journal} {Phys. Rev. E}\ }\textbf {\bibinfo {volume} {102}},\ \bibinfo {pages} {012206} (\bibinfo {year} {2020})}\BibitemShut {NoStop}%
\bibitem [{\citenamefont {Ariaratnam}\ and\ \citenamefont {Strogatz}(2001)}]{Ariaratnam2001}%
  \BibitemOpen
  \bibfield  {author} {\bibinfo {author} {\bibfnamefont {J.~T.}\ \bibnamefont {Ariaratnam}}\ and\ \bibinfo {author} {\bibfnamefont {S.~H.}\ \bibnamefont {Strogatz}},\ }\href@noop {} {\bibfield  {journal} {\bibinfo  {journal} {Phys. Rev. Lett.}\ }\textbf {\bibinfo {volume} {86}},\ \bibinfo {pages} {4278} (\bibinfo {year} {2001})}\BibitemShut {NoStop}%
\bibitem [{\citenamefont {Watanabe}\ and\ \citenamefont {Strogatz}(1993)}]{WS.PRL1993}%
  \BibitemOpen
  \bibfield  {author} {\bibinfo {author} {\bibfnamefont {S.}~\bibnamefont {Watanabe}}\ and\ \bibinfo {author} {\bibfnamefont {S.~H.}\ \bibnamefont {Strogatz}},\ }\href {https://doi.org/10.1103/PhysRevLett.70.2391} {\bibfield  {journal} {\bibinfo  {journal} {Phys. Rev. Lett.}\ }\textbf {\bibinfo {volume} {70}},\ \bibinfo {pages} {2391} (\bibinfo {year} {1993})}\BibitemShut {NoStop}%
\bibitem [{\citenamefont {Watanabe}\ and\ \citenamefont {Strogatz}(1994)}]{WS.PhysD1994}%
  \BibitemOpen
  \bibfield  {author} {\bibinfo {author} {\bibfnamefont {S.}~\bibnamefont {Watanabe}}\ and\ \bibinfo {author} {\bibfnamefont {S.~H.}\ \bibnamefont {Strogatz}},\ }\href {https://doi.org/https://doi.org/10.1016/0167-2789(94)90196-1} {\bibfield  {journal} {\bibinfo  {journal} {Physica D: Nonlinear Phenomena}\ }\textbf {\bibinfo {volume} {74}},\ \bibinfo {pages} {197} (\bibinfo {year} {1994})}\BibitemShut {NoStop}%
\bibitem [{\citenamefont {Pikovsky}\ and\ \citenamefont {Rosenblum}(2010)}]{pikovsky2010partiallyintegrabledynamicsensembles}%
  \BibitemOpen
  \bibfield  {author} {\bibinfo {author} {\bibfnamefont {A.}~\bibnamefont {Pikovsky}}\ and\ \bibinfo {author} {\bibfnamefont {M.}~\bibnamefont {Rosenblum}},\ }\href {https://doi.org/https://doi.org/10.48550/arXiv.1001.1299} {\bibinfo {title} {Partially integrable dynamics of ensembles of nonidentical oscillators}} (\bibinfo {year} {2010}),\ \Eprint {https://arxiv.org/abs/1001.1299} {arXiv:1001.1299 [nlin.AO]} \BibitemShut {NoStop}%
\bibitem [{\citenamefont {Manoranjani}\ \emph {et~al.}(2021)\citenamefont {Manoranjani}, \citenamefont {Gupta},\ and\ \citenamefont {Chandrasekar}}]{Manoranjani2021}%
  \BibitemOpen
  \bibfield  {author} {\bibinfo {author} {\bibfnamefont {M.}~\bibnamefont {Manoranjani}}, \bibinfo {author} {\bibfnamefont {S.}~\bibnamefont {Gupta}},\ and\ \bibinfo {author} {\bibfnamefont {V.~K.}\ \bibnamefont {Chandrasekar}},\ }\href {https://doi.org/10.1063/5.0055664} {\bibfield  {journal} {\bibinfo  {journal} {Chaos: An Interdisciplinary Journal of Nonlinear Science}\ }\textbf {\bibinfo {volume} {31}},\ \bibinfo {pages} {083130} (\bibinfo {year} {2021})},\ \Eprint {https://arxiv.org/abs/https://pubs.aip.org/aip/cha/article-pdf/doi/10.1063/5.0055664/14636370/083130\_1\_online.pdf} {https://pubs.aip.org/aip/cha/article-pdf/doi/10.1063/5.0055664/14636370/083130\_1\_online.pdf} \BibitemShut {NoStop}%
\bibitem [{\citenamefont {Moyal}\ \emph {et~al.}(2024)\citenamefont {Moyal}, \citenamefont {Rajwani}, \citenamefont {Dutta},\ and\ \citenamefont {Jalan}}]{Jalan_PRE2024}%
  \BibitemOpen
  \bibfield  {author} {\bibinfo {author} {\bibfnamefont {B.}~\bibnamefont {Moyal}}, \bibinfo {author} {\bibfnamefont {P.}~\bibnamefont {Rajwani}}, \bibinfo {author} {\bibfnamefont {S.}~\bibnamefont {Dutta}},\ and\ \bibinfo {author} {\bibfnamefont {S.}~\bibnamefont {Jalan}},\ }\href {https://doi.org/10.1103/PhysRevE.109.034211} {\bibfield  {journal} {\bibinfo  {journal} {Phys. Rev. E}\ }\textbf {\bibinfo {volume} {109}},\ \bibinfo {pages} {034211} (\bibinfo {year} {2024})}\BibitemShut {NoStop}%
\bibitem [{\citenamefont {Rohr}\ \emph {et~al.}(2019)\citenamefont {Rohr}, \citenamefont {Berner}, \citenamefont {Lameu}, \citenamefont {Popovych},\ and\ \citenamefont {Yanchuk}}]{Rohr2019}%
  \BibitemOpen
  \bibfield  {author} {\bibinfo {author} {\bibfnamefont {V.}~\bibnamefont {Rohr}}, \bibinfo {author} {\bibfnamefont {R.}~\bibnamefont {Berner}}, \bibinfo {author} {\bibfnamefont {E.~L.}\ \bibnamefont {Lameu}}, \bibinfo {author} {\bibfnamefont {O.~V.}\ \bibnamefont {Popovych}},\ and\ \bibinfo {author} {\bibfnamefont {S.}~\bibnamefont {Yanchuk}},\ }\href {https://doi.org/10.1371/journal.pone.0225094} {\bibfield  {journal} {\bibinfo  {journal} {PLoS ONE}\ }\textbf {\bibinfo {volume} {14}},\ \bibinfo {pages} {e0225094} (\bibinfo {year} {2019})}\BibitemShut {NoStop}%
\end{thebibliography}%

\end{document}